\documentclass[prl,superscriptaddress,twocolumn]{revtex4}
\usepackage[paperwidth=210mm,paperheight=297mm,centering,hmargin=2cm,vmargin=2cm]{geometry}
\usepackage[utf8]{inputenc}
\usepackage{amsmath}
\usepackage{amsfonts}
\usepackage{amssymb}
\usepackage{graphicx}
\usepackage{braket}
\usepackage{dsfont}
\usepackage{framed}
\usepackage{float}
\usepackage{physics}
\usepackage{epstopdf}
\usepackage{physics}
\usepackage{multirow}
\usepackage{amsthm}
\usepackage{color}
\usepackage{tabularx}
\usepackage{algorithm}
\floatname{algorithm}{Protocol}
\usepackage{algorithmic}
\usepackage{mathtools}
\usepackage{appendix}
\usepackage{array}
\usepackage{dsfont}
\usepackage{array}
\usepackage{tabularx}
\usepackage{listings}
\usepackage{makecell}
\usepackage{setspace}
\usepackage{xcolor}
\usepackage[mathscr]{eucal}

\newenvironment{proof-sketch}{%
\proof}{\endproof}

\newcommand\Mycomb[2][^n]{\prescript{#1\mkern-0.5mu}{}C_{#2}}

\newtheorem{manualtheoreminner}{Theorem}
\newenvironment{manualtheorem}[1]{%
  \renewcommand\themanualtheoreminner{#1}%
  \manualtheoreminner
}{\endmanualtheoreminner}

\newtheorem{manuallemmainner}{Lemma}
\newenvironment{manuallemma}[1]{%
  \renewcommand\themanuallemmainner{#1}%
  \manuallemmainner
}{\endmanuallemmainner}

\newcounter{protocol}

\newcommand{\zbold}{\overline{\textbf{z}}}
\newcommand{\xbold}{\overline{\textbf{x}}}

\newcommand{\pbold}{\overline{\textbf{p}}}

\newcommand{\wbold}{\overline{\textbf{w}}}
\newcommand{\vbold}{\overline{\textbf{v}}}

\usepackage{hyperref}
\hypersetup{
    colorlinks=true, 
    linkcolor=red, 
    linktoc=all 
}

\usepackage{color}

\begin{document}
\title{Verification of graph states in an untrusted network}
\author{Anupama Unnikrishnan}
\affiliation{Clarendon Laboratory, University of Oxford, Parks Road, Oxford OX1 3PU, UK}
\author{Damian Markham}
\affiliation{LIP6, CNRS, Sorbonne Universit\'e, 4 Place Jussieu, 75005 Paris, France}
\affiliation{JFLI, CNRS / National Institute of Informatics, University of Tokyo, Tokyo, Japan}

\begin{abstract}
Graph states are a large class of multipartite entangled quantum states that form the basis of schemes for quantum computation, communication, error correction, metrology, and more. In this work, we consider verification of graph states generated by an untrusted source and shared between a network of possibly dishonest parties. This has implications in certifying the application of graph states for various distributed tasks. 
We present a protocol which is globally efficient for a large family of useful graph states, including cluster states, GHZ states, cycle graph states and more. For general graph states, efficiency with respect to the security parameter is maintained, though there is a cost increase with the size of the graph state. The protocols are practical, requiring only multiple copies of the graph state, local measurements and classical communication.

\end{abstract}

\maketitle
The usefulness of graph states extends across the field of quantum information, with their inherent multipartite entanglement leading them to be promising candidates for a variety of tasks. 
As quantum states that correspond to mathematical graphs, where vertices represent qubits and edges preparation entanglement, graph states may possess varied entanglement that render them ideal for applications ranging from quantum computation \cite{Raussendorf2001, Fitzsimons2017, Kashefi2017} and error correction \cite{Gottesman1997, Bella} to cryptography \cite{Markham, Unnikrishnan2019}, metrology \cite{Toth2014, Friis2017, Shettell2019} and more \cite{Hein2004, Hein2006, markham2018quantum}.

As such, the verification of these resource states is a natural problem, and has been the topic of much investigation, under different assumptions of trust and particular questions of interest \cite{eisert2019quantum}. For example, in tomography, one wishes to find the density matrix of the state, whilst trusting measuring devices and that the state is not correlated across preparations, \emph{i.e.} that it is independent and identically distributed (i.i.d.) \cite{DAriano2003}. With entanglement witnesses, on the other hand, we have the same trust settings but are interested only in demonstrating entanglement \cite{Jungnitsch2011}.
In cryptographic settings, one is often interested in the case where measurement devices are trusted, but the sources of the states are not, and in fact not even that they are i.i.d. \cite{Markham2018, Hayashi2015, Takeuchi2018, Takeuchi2019, zhu2019efficient}.
Going further, in self-testing and device-independent approaches, one does not trust the source or the measurement devices \cite{McKague2013}. These techniques have had applications across quantum information for verification of delegated quantum computation \cite{Markham2018, Hayashi2015, Takeuchi2018, Takeuchi2019}, simulation \cite{Bermejo-Vega2018}, sampling and network tasks \cite{Markham2018, Bell}.

In all these examples, however, all of the participating parties behave in an honest way; it is the source or their devices which are not trusted.
One may also be interested in a stronger type of scenario, where a graph state is distributed over a network of  parties each holding a share of the state, but where, crucially, not all the parties behave in an honest way.
This is an important cryptographic scenario, which has applications, for example, in the task of secure multiparty quantum computation \cite{Crepeau2002, Ben-Or2008}, involving a set of possibly dishonest parties who wish to compute some function of their inputs. The first efforts to verify states in this scenario was for GHZ states \cite{Pappa2012, McCutcheon2016, yehia2020composable}. This has already found application, for example, in anonymous transmission of quantum messages \cite{Unnikrishnan2019}. The ubiquity of graph states for quantum information naturally suggests the interest in generalising such an approach to all graph states.

In this work, we present a way of verifying graph states across a network with untrusted parties. 
Our protocol is globally efficient in the number of copies required for families of graph states, including many graph states of interest. These include the complete graph state (locally equivalent to the GHZ state), one- and two-dimensional cluster states, and cyclic graph states, which are resource states for quantum metrology \cite{Toth2014, Shettell2019}, universal quantum computation \cite{Raussendorf2001, Raussendorf2003} and error correction \cite{Gottesman1997, Bella} respectively.
For an arbitrary graph state, our protocol maintains efficiency with respect to the security parameter (\emph{i.e.} how sure we are that the state is good), but requires a number of copies of the graph state that grows exponentially with the size of the graph state. This cost arises since, in general, we lack certain symmetries which allow for efficiency. For small or fixed-size networks requiring fixed-size graph states, this cost may not be too burdensome, as there the security parameter efficiency dominates.

For ease of presentation, we break our work down into two protocols. The first follows very closely to \cite{Takeuchi2019}, extending its use to dishonest networks, but requires an assumption on one party (the Verifier) to act honestly. 
Building on this, the second protocol is more involved, but removes this assumption. In this way, all parties are equally treated with no assumption of their honest behaviour, and we refer to this as the symmetric protocol.

\textit{Communication scenario.---} 
Our network consists of $n$ parties, split into sets of honest and dishonest parties, which we will denote by $H$ and $D$, respectively. 
An untrusted source is asked to produce the $n$-qubit graph states requested by the parties, but may provide any states they wish, including entangled across copies and with possible registers reserved by the source for later use.
The parties are only required to apply local operations and measurements, and communicate classically. 
While the honest parties follow the protocol, the dishonest parties might not, and can work together and do anything to their part of the state, as well as collaborate with the source to disrupt the protocol. 
Note that the honest parties do not need to know the identity of other honest parties, whereas the dishonest parties are assumed to have all the information about who is honest and dishonest. 
Each pair of parties must share a private classical channel. For the symmetric version of the protocol, we also require a trusted common random source.

\textit{Protocol.---}
Before we begin, let us give a brief description of stabilisers. 
The $n$-qubit graph state $\ket{\mathscr{G}}$ associated to a graph $\mathscr{G}$ can be uniquely specified by its stabiliser group $S$ as follows.
To each qubit $i \in \{1, ..., n \}$, we associate a \emph{stabiliser generator} $K_i = X_i \prod_{e \in N(i)} Z_e$, where $N(i)$ is the set of neighbouring vertices to $i$ in the graph and $X_i$ and $Z_i$ are the Pauli operators acting on qubit $i$.
The full \emph{stabiliser group} $S$ is generated by taking all $2^n$ products of the stabiliser generators. The graph state $\ket{\mathscr{G}}$ is the unique $+1$ eigenstate of every stabiliser $S_j\in S$.

An intuitive way to test a particular graph state is then to check that all stabiliser measurements give a $+1$ outcome. If all parties are honest, this can only happen if the source has sent them the ideal graph state (since it is the unique $+1$ eigenstate). By asking for many copies and randomly choosing which stabiliser to measure after the source has sent each copy, the parties can prevent the source from creating some other state that would pass an individual stabiliser measurement but not the full set. 
In fact, this is the essential idea behind many of the verification protocols for graph states, with details differing depending on trust, security statements and efficiency \cite{markham2015practical,Markham2018, Hayashi2015, Takeuchi2018, Takeuchi2019,zhu2019efficient,zhu2019efficienthyper,pallister2018optimal}.
Here, we extend this approach to account for any number of dishonest parties, who may work together and coordinate with the dishonest source.

To achieve this goal, we will use and adapt the verification protocol from \cite{Takeuchi2019}. 
Our  Protocol \ref{alg:takeuchi} is almost identical to that of \cite{Takeuchi2019}, although the security proof requires completely new elements at the state level to deal with dishonest parties.
At this stage however, the protocol still requires one party, the Verifier, to behave honestly. 
We then adapt it to build a new protocol to deal with the possibility of dishonest Verifiers, which we call the symmetric scenario (\emph{i.e.} all parties are treated equally), in  Protocol \ref{alg:dishonestver}.

The presence of dishonest parties affects the efficiency of the protocol. While the protocol in \cite{Takeuchi2019} only requires measuring the stabiliser generators, in order to extend this to any graph state with any number of dishonest parties, the parties must measure the full set of stabilisers, which leads to the requirement of exponentially many copies with graph size. This scaling is due to the power of our adversarial model. We will show afterwards how to reduce this requirement if we have more information about the dishonest parties, or for specific families of graph states of interest. 

\begin{algorithm}[t]
\caption{\textsc{Verification of graph state}}
\begin{flushleft}
\textit{Input}: The parties choose values of $N_{test}, N_{total}$. Let $\mathcal{S}$ be the set of test measurements, and $\mathsf{J} = \abs{\mathcal{S}}$.  \\

\end{flushleft}
\begin{algorithmic}[1]
\STATE An untrusted source generates $N_{total}$ copies of the graph state, and sends the shares to the parties. \\ \ 
\STATE The parties repeat the following for $j = 1, ..., \mathsf{J}$: 
\begin{enumerate}
\item[(a)] The Verifier chooses $N_{test}$ copies from the remaining $N_{total} - (j-1) N_{test}$ copies independently and uniformly at random.
\item[(b)] For each copy, the Verifier instructs each party to perform the measurement corresponding to their part of the stabiliser $\mathcal{S}_j$.
\item[(c)] For each copy, the parties send their measurement outcome to the Verifier, who calculates the total measurement outcome. The copy passes the test if the total measurement outcome is $+1$. Let $N_{pass, j}$ be the number of copies that pass the stabiliser test for $\mathcal{S}_j$. 
\end{enumerate}
\STATE The Verifier uniformly randomly chooses a single copy from the remaining $\mathsf{N} \equiv N_{total} - \mathsf{J} N_{test}$ copies that were not used for the tests in the previous steps. The chosen single copy is called the target copy. The others are discarded. \\ \
\STATE If $N_{pass} \equiv \sum_{j = 1}^{\mathsf{J}} N_{pass, j} \geq \mathsf{J} N_{test} - \frac{N_{test}}{2  \mathsf{J}}$, the parties use the target copy for their application; otherwise, the target copy is discarded. 
\end{algorithmic}
\label{alg:takeuchi}
\end{algorithm}

\textit{Security analysis.---}
We first consider the case of an arbitrary $n$-qubit graph state, denoted by $\ket{\mathscr{G}}$, and assume an honest Verifier. 
The parties perform Protocol \ref{alg:takeuchi}, measuring the full set of $\mathsf{J}=2^n$ stabilisers (\emph{i.e.}  we take the set of test measurements, $\mathcal{S}$, to be the full stabiliser group $S$), to test whether the state they receive in each round, $\ket{\Psi}$, is the ideal graph state $\ket{\mathscr{G}}$, even in the presence of dishonest parties. 

In a network with dishonest parties who may collaborate with the source of the state, we may take the state $\ket{\Psi}$ to be pure, as the dishonest parties can purify it by adding a reference system. Further, in such a network, we can only make statements on the fidelity up to local  operations on the dishonest part. By Uhlmann's Theorem \cite{Uhlmann1976}, this refers to the fidelity between the reduced states of the honest parties.

Let us then take $\rho_H^{\ket{\mathscr{G}}}$ to be the reduced state of the honest parties in the ideal case (\emph{i.e.} when they share $\ket{\mathscr{G}}$), and $\rho_H^{avg}$ to be the reduced state of the honest parties of the averaged state of the 
target copy 
(over all random selections from the remaining $\mathsf{N}$ copies). 
We prove the security of our protocol in the following Theorem for the case of an honest Verifier.

\begin{manualtheorem}{1}
Assuming an honest Verifier, if we set $N_{total} = 2 \times 2^n N_{test}$, $N_{test} = \lceil{ m 2^{4n} \ln 2^n}\rceil$, 
and $\mathcal{S}$ as the full set of stabilisers, the probability that the fidelity of the averaged state of 
the target copy (over all possible choices of the tested copies and target copy)
in Protocol \ref{alg:takeuchi} satisfies
\begin{align}
F(\rho_H^{\ket{\mathscr{G}}}, \rho_H^{avg}) 
\geq 1 - \frac{2\sqrt{c}}{2^n} - 2 \times 2^n \Big( 1 - \frac{N_{pass}}{2^n N_{test}} \Big)
\end{align}
is at least $1 - (2^n)^{1-\frac{2cm}{3}}$, 
where
$m, c$ are positive constants chosen such that $\frac{3}{2m} < c <  \frac{(2^n-1)^2}{4}$.
\label{th:graphbigproof}
\end{manualtheorem}

\begin{proof-sketch}  \renewcommand{\qedsymbol}{}
Our proof has three main stages. In the first stage, we show that if all stabiliser tests pass perfectly, the state in each round of the protocol must be $\ket{\Psi} = U_D \ket{\mathscr{G}}$, where $U_D$ is a unitary on the dishonest part of the state. This is the main part of our proof.
To do this, we demonstrate that, for the Schmidt decomposition of the state corresponding to the partition $(H, D)$, the Schmidt basis of the honest parties is the same for the actual and ideal states. We illustrate this by writing the unknown actual state as 
\begin{align}
\ket{\Psi} & = \sum_{\zbold} \alpha_{\zbold} \ket{\zbold}_H \otimes \ket{\psi_{\zbold}}_D, \label{eq:first}\\
\ket{\Psi} & = \sum_{\xbold} \beta_{\xbold} \ket{\mathscr{H}_{(\xbold)}}_H \otimes \ket{\phi_{\xbold}}_D,
\label{eq:second}
\end{align}
where $\zbold, \xbold$ are classical $|H|$-bit strings, $\alpha_{\zbold}, \beta_{\xbold}$ are unknown complex coefficients, and $\{ \ket{\zbold} \}, \{ \ket{\mathscr{H}_{(\xbold)}} \}$ form complete bases for the honest side, where $\ket{\mathscr{H}_{(\xbold)}}$ is the state corresponding to the honest subgraph with $\sigma_Z$s applied on vertices according to $\xbold$ (that is, $\sigma_Z$ applied to the qubit of party $i$ iff $x_i=1$), and $\ket{\psi_{\zbold}}, \ket{\phi_{\xbold}}$ are the corresponding states on the dishonest side.
We identify, purely from the fact that all stabiliser generator tests pass perfectly, which elements in $\{ \ket{\psi_{\zbold}} \}$ and $\{ \ket{\phi_{\xbold}} \}$ are orthogonal to each other. Passing the full set of stabiliser tests allows us to see which $\beta_{\xbold}$ must be zero; this tells us which terms of $\ket{\zbold}$ must correspond to the same $\alpha_{\zbold}, \ket{\psi_{\zbold}}$. 
We then show that each unique $\alpha_{\zbold} = \pm \frac{1}{\sqrt{2^{|H|}}}$. Substituting this information in Equation (\ref{eq:first}), we see that the honest Schmidt basis of $\ket{\Psi}$ must be equal to the honest Schmidt basis of $\ket{\mathscr{G}}$.
Thus, the only state $\ket{\Psi}$ that can pass every stabiliser test perfectly is such that $\rho_H^{\ket{\Psi}} = \rho_H^{\ket{\mathscr{G}}}$ (or equivalently, $\ket{\Psi} = U_D \ket{\mathscr{G}}$). 

In the second stage, we employ the Serfling bound to find that, after the $2^n$ test measurements, 
the fraction of states out of the remaining untested copies that 
would pass all tests perfectly
is at least 
$1 - \frac{2\sqrt{c}}{2^n} - 2 \times 2^n \big( 1 - \frac{N_{pass}}{2^n N_{test}} \big)$, 
with probability at least $1 - (2^n)^{1-\frac{2cm}{3}}$. 
This follows straightforwardly using the method of \cite{Takeuchi2019}, but with $2^n$ test measurements corresponding to the full set of stabilisers.

In the final stage, we translate this statement to the reduced state of the honest parties of 
the target copy,
which gives our final result.

The full proof is given in Appendix I.
\end{proof-sketch}

\

\begin{figure*}[t]
\centering
\includegraphics[trim = 2cm 21cm 1cm 0cm, width=0.9\textwidth]{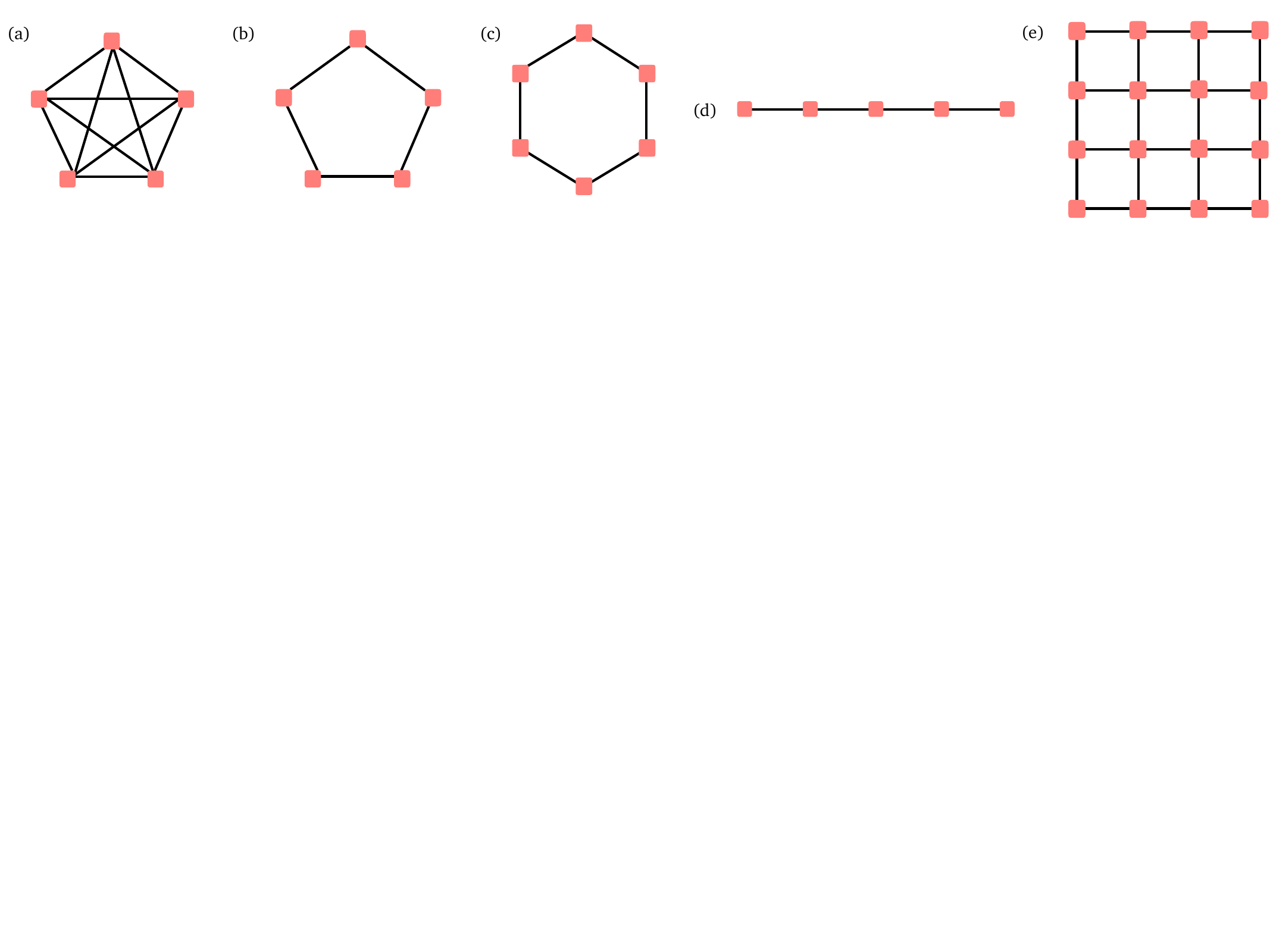}
\caption{Examples of a (a) complete graph, (b) pentagon graph, (c) cycle graph, (d) $\mathsf{1D}$ cluster state, and (e) $\mathsf{2D}$ cluster state. For quantum states that correspond to these types of graphs, one can perform efficient verification using Protocol \ref{alg:takeuchi}, possibly with some information regarding the dishonest parties, as shown in Theorem \ref{th:allexamples}.}
\label{fig:examples}
\end{figure*}

\textit{Efficient examples.---}
For certain graph states or sets of dishonest parties, we can reduce the resources required for verification; in fact, we see that simply by measuring the stabiliser generators, the desired statement can be made about the fidelity of the unknown state. 
Then, the parties can instead run Protocol \ref{alg:takeuchi} with a simplified set of test measurements (where $\mathsf{J} = n$). Such examples are given in the following Theorem, whose graph states are shown in Figure \ref{fig:examples}.


\begin{manualtheorem}{2}
If $\ket{\mathscr{G}}$ is either
\begin{enumerate}
\item[A.] a complete graph state with dishonest parties anywhere in the network, or
\item[B.] a pentagon graph state with either one, three or four dishonest parties anywhere in the network, or two dishonest parties who are adjacent, or 
\item[C.] a cycle graph state with either one, $n-2$ or $n-1$ dishonest parties anywhere in the network, or any other number of dishonest parties who are adjacent, or 
\item[D.] a $\mathsf{1D}$ cluster state with either one or $n-1$ dishonest parties anywhere in the network, or any other number of adjacent honest or dishonest parties, or 
\item[E.] a $\mathsf{2D}$ cluster state with either one or $n-1$ dishonest parties anywhere in the network, or any other set of adjacent dishonest parties that forms a square or rectangle anywhere in the network,
\end{enumerate}
and we set $N_{total} = 2n N_{test}, N_{test} = \lceil{ m n^4 \ln n}\rceil$, 
and $\mathcal{S}$ as the set of stabiliser generators, assuming an honest Verifier, 
the probability that the fidelity of the averaged state of 
the target copy (over all possible choices of the tested copies and target copy) 
in Protocol \ref{alg:takeuchi} satisfies
\begin{align}
F(\rho_H^{\ket{\mathscr{G}}}, \rho_H^{avg}) 
\geq 1 - \frac{2\sqrt{c}}{n} - 2n \Big( 1 - \frac{N_{pass}}{n N_{test}} \Big) 
\end{align}
is at least $1 - n^{1 - \frac{2cm}{3}}$, where 
$m, c$ are positive constants chosen such that $\frac{3}{2m} < c < \frac{(n-1)^2}{4}$.
\label{th:allexamples}
\end{manualtheorem}
\begin{proof-sketch}  \renewcommand{\qedsymbol}{}
First, we see that, for the examples A-E in Theorem 2, testing the stabiliser generators is sufficient to identify the terms that are zero in Equation (\ref{eq:second}). 
Then, we follow the remainder of the proof of Theorem \ref{th:graphbigproof}, but with $n$ test measurements corresponding to the generator tests, and extend our results to 
the target copy
in the final step.
The full details are given in Appendix II.

\end{proof-sketch}

Noting that the GHZ state is locally equivalent to the complete graph state, example A offers an alternative to the verification protocol in \cite{Pappa2012}.

\textit{Symmetric protocol.---}
Up to now, we have assumed that the party acting as the Verifier is honest. We will now remove this assumption. To do so, we increase the number of tested copies from $N_{test}$ tests per stabiliser to $\lambda N_{test}$, and make use of a trusted common random source (CRS) that provides the parties with shared randomness (following \cite{Pappa2012}). 
The CRS is used to replace the Verifier's choices and to pick which party acts as the Verifier. In this way, the honest parties are protected against a dishonest Verifier who may attempt to cheat.
While using a CRS is an additional resource, it is natural that it is also required, as any protocol able to verify an entangled state can use this state to generate a CRS (for example, by sharing a GHZ state and measuring in the computational basis). A CRS can be created, for example, if a third or more of the network is honest, using \cite{chaum1988multiparty}.

Such a symmetric protocol is given in Protocol \ref{alg:dishonestver}, with the following adapted Theorem, whose proof is given in Appendix III.

\begin{manualtheorem}{3}
If we set $N_{total} = (\lambda+1)\lambda \mathsf{J} N_{test}$ and $N_{test} = \lceil{ m \mathsf{J}^4 \ln \mathsf{J} }\rceil$, 
the probability that the fidelity of the averaged state of 
the target copy (over all possible choices of the tested copies and target copy)
in Protocol \ref{alg:dishonestver} satisfies
\begin{align}
F(\rho_H^{\ket{\mathscr{G}}}, \rho_H^{avg}) 
 \geq & 1 - \Big( \frac{1}{\lambda} - \frac{1}{\lambda^2} \Big) \nonumber \\
& \ - \Big( 1 + \frac{1}{\lambda} \Big) \Big[ \frac{\sqrt{c}}{\mathsf{J}} + \lambda \mathsf{J} - \frac{N_{pass}}{N_{test}} \Big]
\end{align}
is at least $
\Big[ 1 - \sum_{x = 0}^{\lambda} \big( 1 - \frac{\abs{H}}{n} \big)^x \big( \frac{\abs{H}}{n} \mathsf{J}^{-\frac{2cm}{3}} \big)^{\lambda - x} \Big]^{\mathsf{J}},
$
where
$m, c$ are positive constants chosen such that the fidelity and probability expressions are greater than zero.
\label{th:graphdishonestverproof}
\end{manualtheorem}

Replacing $\mathsf{J}$ by $\mathsf{J}=2^n$ for a general graph state, and $\mathsf{J}=n$ for the special cases in Theorem \ref{th:allexamples}, we recover symmetric versions of Theorems \ref{th:graphbigproof} and \ref{th:allexamples}.
The extra parameter $\lambda$ allows the parties to further tailor the protocol to achieve their desired bounds, while taking into account their experimental limitations.

\bigskip

\begin{algorithm}[t]
\caption{\textsc{Symmetric Protocol for Verification of Graph State}}
\begin{flushleft}
\textit{Input}: The parties choose values of $N_{test}, N_{total}, \lambda$. Let $\mathcal{S}$ be the set of test measurements, and $\mathsf{J} = \abs{\mathcal{S}}$.  
\end{flushleft}
\begin{algorithmic}[1]
\STATE An untrusted source generates $N_{total}$ copies of the graph state, and sends the shares to the parties. \\ \ 
\STATE The parties repeat the following for $j = 1, ..., \mathsf{J}$: 
\begin{enumerate}
\item[(a)] The CRS chooses $\lambda N_{test}$ copies from the remaining $N_{total} - (j-1) \lambda N_{test}$ copies independently and uniformly at random. These are split into $\lambda$ sets of $N_{test}$ copies, again at random by the CRS. For each set of $N_{test}$ copies, the CRS chooses a random party to be the Verifier. 
\item[(b)] For each copy, the corresponding Verifier instructs each party to perform the measurement corresponding to their part of the stabiliser $\mathcal{S}_j$.
\item[(c)] For each copy, the parties send their measurement outcome to the Verifier, who calculates the total measurement outcome. The copy passes the test if the total measurement outcome is $+1$. Let $N_{pass, j}$ be the number of copies that pass the stabiliser test for $\mathcal{S}_j$. 
\end{enumerate}
\STATE The CRS uniformly randomly chooses a single copy from the remaining $\mathsf{N} \equiv N_{total} - \lambda \mathsf{J} N_{test}$ copies that were not used for the tests in the previous steps. The chosen single copy is called the target copy. The others are discarded. \\ \
\STATE If $N_{pass} \equiv \sum_{j = 1}^{\mathsf{J}} N_{pass, j} \geq \lambda \mathsf{J} N_{test} - \frac{N_{test}}{2  \mathsf{J}}$, the parties use the target copy for their application; otherwise, the target copy is discarded. 
\end{algorithmic}
\label{alg:dishonestver}
\end{algorithm}

\textit{Discussion.---}
In this work, we have shown how to verify any graph state shared between a network of parties who may or may not be trusted. Our protocols are easily implementable: performing a verified version of some task using graph states simply requires additional copies of the graph state. Indeed, recent experiments verifying the GHZ state \cite{McCutcheon2016} and secret sharing graph states \cite{Bell} for trusted parties would need minimal changes to carry out our protocol, while giving increased security. The results here naturally extend to qudit graphs, as in \cite{Takeuchi2019}. 


We briefly consider the robustness and comparative efficiency of our protocols. 
In terms of robustness, we are interested in how the fidelity statements and probability of accept change when noise is introduced \cite{Takeuchi2019,unnikrishnan2019authenticated}. Our protocols  follow the robustness properties of \cite{Takeuchi2019}, in particular the scaling of the accept probability with the number of tests (see Appendix IV). 
Unsurprisingly, the additional tests required for the general case diminish robustness, whereas if only the generators need to be tested, such as for the examples in Theorem \ref{th:allexamples}, we recover the linear scaling of \cite{Takeuchi2019}. Moving to the symmetric case (Theorem \ref{th:graphdishonestverproof}) does not greatly affect the robustness.

With respect to efficiency, we can compare with the case where all parties are trusted, studied in depth in \cite{pallister2018optimal, zhu2019efficient,zhu2019efficienthyper}. 
In particular, these works also give lower bounds on the number of copies of a graph state needed to certify a given fidelity with a given probability of accept. Since our scenario is more paranoid 
(theirs corresponds to all parties being honest),
these works give lower bounds for our protocols too. 
For the families of interest in Theorem \ref{th:allexamples}, we achieve the same efficiency as \cite{Takeuchi2019}. For general graph states, where we cannot make use of symmetries, we suffer from an exponential overhead with the size of the graph state, but for a fixed-size graph we maintain efficiency with respect to the security (fidelity and probability).
It is likely that further optimisations can be made for efficiency, either by adapting the protocol or different proof techniques \cite{pallister2018optimal, zhu2019efficient, zhu2019efficienthyper,unnikrishnan2019authenticated,hayashi2018self}, though it is not clear the same lower bounds will hold. 

The efficient examples from Theorem \ref{th:allexamples} suggest several applications. Cluster states are universal for quantum computation \cite{Raussendorf2001, Raussendorf2003} and can be used in a distributed setting, meaning their verification can be applied for secure multiparty computation \cite{Crepeau2002, Ben-Or2008}. 
The complete graph state, or locally equivalent GHZ state, is the optimal resource for quantum metrology \cite{Toth2014, Shettell2019}, where verification would have applications for distributed or delegated sensing \cite{huang2019,yin2020,komar2014quantum,ShettellCryptoMetro} in untrusted networks. The cycle graph state is both the most efficient state for error correction and a resource for secret sharing \cite{Marin13}, again, with applications for further security in untrusted networks \cite{Bell}.

\textit{Acknowledgements.---} We acknowledge funding from the EPSRC, the ANR through the ANR-17-CE24-0035 VanQuTe project, and the European Union's Horizon 2020 Research and Innovation Programme under Grant Agreement No. 820445 (QIA).


\onecolumngrid
\appendix
\normalsize 
\section{APPENDIX I: PROOF OF THEOREM \ref{th:graphbigproof}}

Using Protocol \ref{alg:takeuchi}, a network of parties, out of which any number may be dishonest, can verify any shared graph state. We now prove the following Theorem to illustrate the security of this protocol, assuming for now that the Verifier is an honest party.

\begin{manualtheorem}{1}
Assuming an honest Verifier, if we set $N_{total} = 2 \times 2^n N_{test}$, $N_{test} = \lceil{ m 2^{4n} \ln 2^n}\rceil$, and $\mathcal{S}$ as the full set of stabilisers, the probability that the fidelity of the averaged state of
the target copy (over all possible choices of the tested copies and target copy) 
in Protocol \ref{alg:takeuchi} satisfies
\begin{align}
F(\rho_H^{\ket{\mathscr{G}}}, \rho_H^{avg}) 
\geq 1 - \frac{2\sqrt{c}}{2^n} - 2 \times 2^n \Big( 1 - \frac{N_{pass}}{2^n N_{test}} \Big)
\end{align}
is at least $1 - (2^n)^{1-\frac{2cm}{3}}$, 
where
$m, c$ are positive constants chosen such that $\frac{3}{2m} < c <  \frac{(2^n-1)^2}{4}$.
\end{manualtheorem}
\begin{proof}
Our proof proceeds in stages. We will start by proving that if all stabiliser tests pass perfectly, then the state must be the ideal graph state, up to local unitaries on the dishonest parts of the state. This means that in order to pass perfectly, the source must create entanglement between the honest and dishonest parties. This is given in the following Lemma.

\begin{manuallemma}{1.1}
The only state that can pass all stabiliser tests perfectly in each round is $\ket{\Psi} = U_D \ket{\mathscr{G}}$, where $U_D$ is a unitary on the dishonest part of the state. 
\label{lem:windsurf}
\end{manuallemma}
\begin{proof}

Let $s$ be the Schmidt rank of the graph state $\ket{\mathscr{G}}$ corresponding to the partition $(H, D)$ into honest and dishonest vertices, $\ket{\mathscr{H}}$ be the state corresponding to the honest subgraph, and $E_\mathscr{H}$ be the set of edges within the honest subgraph.

Let $bn(x)$ be a function that converts a decimal number $x$ into its binary representation, and let us define classical $|H|$-bit strings $\zbold, \xbold \in \{ bn(0), ..., bn(2^{|H|} -1) \}$, where $|H|$ is the number of honest parties in the network. We will use $z_a$ to refer to the $a^{\text{th}}$ element of $\zbold$.

We will start by writing the Schmidt decomposition of the ideal state for the partition $(H, D)$ from \cite{Hein2004}. Some terms in $\{ \ket{\zbold} \}$ may be grouped together as they have the same dishonest part, as given by
\begin{align}
\ket{\mathscr{G}} 
& = \frac{1}{\sqrt{2^{|H|}}} \sum_{\zbold} (-1)^{\underset{(k, l) \in E_H}{\oplus} z_k \land z_l} \ket{\zbold}_H \otimes  \prod_{a \in H} \big( \prod_{b \in N(a)} \sigma_Z^{(b)} \big)^{z_a} \ket{\mathscr{G-H}}_D \\ 
& = \frac{1}{\sqrt{2^{|H|}}} \sum_{\zbold} (-1)^{\underset{(k, l) \in E_H}{\oplus} z_k \land z_l} \ket{\zbold}_H \otimes \ket{\mathscr{G-H}_{(f(\zbold))}}_D,
\label{eq:schmitty}
\end{align}
where $f(\zbold)$ depends on $\underset{i \in N(d)}{\oplus} z_i, \forall d \in D$, and $\mathscr{G-H}$ is the dishonest subgraph. Then, $\ket{\mathscr{\mathscr{G-H}}_{(f(\zbold))}}$ is the state corresponding to the dishonest subgraph with $\sigma_Z$s applied to some vertices depending on $\zbold$. This means that for terms where $\underset{i \in N(d)}{\oplus} z_i$ is the same for all $d \in D$, the corresponding dishonest part $\ket{\mathscr{\mathscr{G-H}}_{(f(\zbold))}}$ is the same, and so the $\ket{\zbold}$s of these are grouped together. 

Let $\ket{\mathscr{H}_{(\xbold)}}$ be the state corresponding to the honest subgraph with $\sigma_Z$s applied according to $\xbold$ (for example, $\ket{\mathscr{H}_{(001)}}$ corresponds to a 3-qubit graph with $\sigma_Z$ applied to the third qubit). The set $\{ \ket{\mathscr{H}_{(\xbold)}} \}$ for all $\xbold$ forms a complete basis (the graph state basis), as does the set $\{ \ket{\zbold} \}$. The unknown state in any round of Protocol \ref{alg:takeuchi} can then be written using the following decomposition:
\begin{align}
\text{honest computational basis: \ \  } \ket{\Psi} & = \sum_{\zbold} \alpha_{\zbold} \ket{\zbold}_H \otimes \ket{\psi_{\zbold}}_D, \label{eq:compbas} \\
\text{honest subgraph basis:  \ \ } \ket{\Psi} & = \sum_{\xbold} \beta_{\xbold} \ket{\mathscr{H}_{(\xbold)}}_H \otimes \ket{\phi_{\xbold}}_D, \label{eq:subbas}
\end{align}
where $\alpha_{\zbold}, \beta_{\xbold}$ are complex coefficients, and both sets $\{ \ket{\psi_{\zbold}} \}, \{ \ket{\phi_{\xbold}} \}$ are the corresponding states on the dishonest side. So far, we have assumed nothing about the portion of the state in control of the dishonest parties, since no matter what the source supplies, the dishonest parties may do anything to their part to cheat in the protocol.

In Protocol \ref{alg:takeuchi}, the test measurements are the elements $\mathcal{S}_{j \in \{ 1, ..., 2^n \}}$ belonging to the full stabiliser group. These measurements are generated by the stabiliser generators $K_{i \in \{ 1, ..., n \}}$ of each qubit $i$. Thus, the full stabiliser group $\mathcal{S}$ is given by 
\begin{align}
\mathcal{S} = \langle \{ K_i = X_i \prod_{e \in N(i)} Z_e \}_{i=1}^n \rangle. 
\end{align}
We want to determine the form of the state given that all the test measurements pass perfectly. 
Let us now group the test measurements into two sets, such that Group 1 contains only $\mathds{1}, Z$ for the honest part, and Group 2 contains everything else. 

We first consider the measurements in Group 1. 
This set contains the stabiliser generators corresponding to the dishonest qubits, \emph{i.e.} $K_{i \in D}$. 
For each dishonest qubit, the honest part of the corresponding stabiliser generator measurement  will be composed of $Z$ measurements on the honest qubits that are in the neighbourhood of the dishonest qubit, and $\mathds{1}$ on those that are not.
We can then write the set of stabiliser generators belonging to Group 1 as 
\begin{align}
\{ \forall d \in D, \Big[ \underset{i \in N(d)}{\otimes} Z_i \underset{i \notin N(d)}{\otimes} \mathds{1}_i \Big]_H \otimes (M'_d)_D \},
\end{align}
where $(M'_d)_D$ is the measurement on the dishonest part, the form of which does not matter here.
Let $\zbold, \zbold'$ be two strings in the computational basis expansion of $\ket{\Psi}$. 
If $\underset{i \in N(d)}{\oplus} z_i \neq \underset{i \in N(d)}{\oplus} z'_i$, then each honest measurement $\underset{i \in N(d)}{\otimes} Z_i \underset{i \notin N(d)}{\otimes} \mathds{1}_i$ will give different outcomes for $\zbold, \zbold'$, since the outcome of the $Z$ measurement on all $i \in N(d)$ depends on the parity of the string (how many 1s). So, to pass perfectly, $(M'_d)_D$ must also give different outcomes. 
In order to do this, the dishonest parties must be able to guess the honest parties' outcome perfectly, which means that they must be able to perfectly discriminate between the corresponding $\ket{\psi_{\zbold}}, \ket{\psi_{\zbold'}}$; this tells them the outcome they should get in order to pass the test. This means that we must have $\bra{\psi_{\zbold}}\ket{\psi_{\zbold'}} = 0$.
%
%
%
In general, we can write this as 
\begin{align}
 \forall  \zbold, \zbold',   \text{ if } \exists d \in D \text{ such that } \underset{i \in N(d)}{\oplus} z_i \neq \underset{i \in N(d)}{\oplus} z'_i, \text{ then }  \bra{\psi_{\zbold}}\ket{\psi_{\zbold'}} = 0.
 \label{eq:psiorth}
\end{align}

We will now see that if the measurements in Group 2 pass perfectly, we get certain $\beta_{\xbold} = 0$, and for all $\beta_{\xbold} \neq 0$, the corresponding $\ket{\phi_{\xbold}}$s will be orthogonal to each other.

To prove orthogonality, note that the set of stabiliser generators in Group 2 contain all the stabiliser measurements with $X$ in the honest part (the stabiliser generator corresponding to each qubit in the honest set). So, the honest part of the measurements in the group is of the form $\{ X_1 M_2 M_3 ...M_n , M_1 X_2 M_3 ...M_n , M_1 M_2 X_3...M_n , ..., M_1 ... X_n \}$, where the honest measurement $M_{i \in \{1, ..., n\}}$ is either $\mathds{1}, Z$. Since no two states in the $\ket{\mathscr{H}_{(\xbold)}}$ basis will give the same outcome for every one of these measurements, in order to pass perfectly, the dishonest parties must be able to discriminate between all corresponding $\ket{\phi_{\xbold}}$ to perfectly guess the honest outcome. We can write this as
\begin{align}
\forall \xbold \neq \xbold', \bra{\phi_{\xbold}}\ket{\phi_{\xbold'}} = 0.
\label{eq:phiorth}
\end{align}

To determine which $\beta_{\xbold} = 0$, we will use the measurements with dishonest part $M_D' = (\mathds{1}...\mathds{1})'$. Let $A \subset H$ be a set containing vertices in $H$ such that all $d \in D$ has an even number of neighbours in $A$. (Note that there may be multiple such sets $A$.) Now, let $K_i$ be the stabiliser generator of qubit $i$ of $\ket{\mathscr{G}}$, and $k_i$ be the stabiliser generator of qubit $i$ of 
$\ket{\mathscr{H}}$. We can then write a subset of the Group 2 measurements as
\begin{align}
\prod_{i \in A} K_i = \prod_{i \in A} (k_i)_H \otimes \mathds{1}_D'.
\end{align}
When the dishonest parties are asked to measure $\mathds{1}$, they must always give outcome +1. So, in order to pass perfectly, the honest part must also give +1 outcome. 
In the expansion of the state in Equation (\ref{eq:subbas}),
if for any set $A$, we do not have an even number of $\sigma_Z$s on the qubits $i \in A$ in the honest subgraph, then the honest measurement will not give outcome $+1$.
Since the positions of the $\sigma_Z$s are determined by the string $\xbold$, passing the Group 2 measurements perfectly tells us that 
\begin{align}
\text{if } \exists A \text{ s.t. } \underset{i \in A}{\oplus} x_i = 1, & \text{ then } \beta_{\xbold} = 0. 
\label{eq:condition}
\end{align}
In this way, we are left with $s$ non-zero $\beta_{\xbold}$s. 
We can write the set $A$ as 
\begin{align}
A = \{ i \in H \text{ } | \ \exists \zbold, \zbold' \text{ such that } \forall d \in D, \underset{i \in N(d)}{\oplus} z_i = \underset{i \in N(d)}{\oplus} z_i', \text{ and } z_i \oplus z_i' = 1 \}.
\label{eq:a}
\end{align}
Writing $i \in A$ using Equation (\ref{eq:a}), the condition on $\beta_{\xbold}$ in Equation (\ref{eq:condition}) can be written as
\begin{align}
\text{if } \exists & \zbold, \zbold' \text{ such that } \forall d \in D, 
\underset{i \in N(d)}{\oplus} z_i = \underset{i \in N(d)}{\oplus} z_i', \nonumber \\
& \text{then }  \beta_{\xbold} = 0 \text{ if } \underset{i \in H}{\oplus} (z_i \oplus z_i' ) \land x_i = 1.
\label{eq:bxz}
\end{align}
We will now relate the two expressions we have for the honest parts of the state.
The graph state basis can be expressed in terms of the computational basis by 
\begin{align}
\ket{\mathscr{H}_{(\xbold)}} & = \frac{1}{\sqrt{2^{|H|}}}  \sum_{\zbold} (-1)^{\underset{(k, l) \in E_H}{\oplus} z_k \land z_l} (-1)^{\underset{i}{\oplus} x_i \land z_i} \ket{\zbold}.
\end{align}
Substituting in Equation (\ref{eq:subbas}) and grouping, we get
\begin{align}
\ket{\Psi} & =  \frac{1}{\sqrt{2^{|H|}}}  \sum_{\xbold} \beta_{\xbold}  \sum_{\zbold} (-1)^{\underset{(k, l) \in E_H}{\oplus} z_k \land z_l} (-1)^{\underset{i}{\oplus} x_i \land z_i} \ket{\zbold}_H  \ket{\phi_{\xbold}}_D  \nonumber \\
& = \sum_{\zbold} (-1)^{\underset{(k, l) \in E_H}{\oplus} z_k \land z_l} \ket{\zbold}_H \frac{1}{\sqrt{2^{|H|}}}  \sum_{\xbold} \beta_{\xbold} (-1)^{\underset{i}{\oplus} x_i \land z_i}  \ket{\phi_{\xbold}}_D. 
\end{align}
Comparing the above expression with Equation (\ref{eq:compbas}) in terms of the coefficients of each $\ket{\zbold}$,
\begin{align}
\alpha_{\zbold} \ket{\psi_{\zbold}} & = (-1)^{\underset{(k, l) \in E_H}{\oplus} z_k \land z_l}  \frac{1}{\sqrt{2^{|H|}}}  \sum_{\xbold} \beta_{\xbold} (-1)^{\underset{i}{\oplus} x_i \land z_i}  \ket{\phi_{\xbold}}, \text{   } \forall \zbold,  \nonumber \\
(-1)^{\underset{(k, l) \in E_H}{\oplus} z_k \land z_l} \alpha_{\zbold} \ket{\psi_{\zbold}} & = \frac{1}{\sqrt{2^{|H|}}} \sum_{\xbold} \beta_{\xbold} (-1)^{\underset{i}{\oplus} x_i \land z_i} \ket{\phi_{\xbold}}, \text{   } \forall \zbold. 
\label{eq:alpha}
\end{align}
On the other hand, expressing the computational basis in terms of the graph state basis gives
\begin{align}
\ket{\zbold} = \frac{1}{\sqrt{2^{|H|}}} (-1)^{\underset{(k, l) \in E_H}{\oplus} z_k \land z_l} \sum_{\xbold} (-1)^{\underset{i}{\oplus} x_i \land z_i} \ket{\mathscr{H}_{(\xbold)}}.
\end{align}
Substituting in Equation (\ref{eq:compbas}) and grouping, we get 
\begin{align}
\ket{\Psi} & = \frac{1}{\sqrt{2^{|H|}}} \sum_{\zbold} \alpha_{\zbold} (-1)^{\underset{(k, l) \in E_H}{\oplus} z_k \land z_l} \sum_{\xbold} (-1)^{\underset{i}{\oplus} x_i \land z_i} \ket{\mathscr{H}_{(\xbold)}}_H \ket{\psi_{\zbold}}_D  \nonumber \\
 & = \sum_{\xbold} \ket{\mathscr{H}_{(\xbold)}}_H \frac{1}{\sqrt{2^{|H|}}}  \sum_{\zbold} (-1)^{\underset{(k, l) \in E_H}{\oplus} z_k \land z_l} (-1)^{\underset{i}{\oplus} x_i \land z_i} \alpha_{\zbold} \ket{\psi_{\zbold}}_D. 
\end{align}
Comparing the above expression with Equation (\ref{eq:subbas}) in terms of the coefficients of $\ket{\mathscr{H}_{(\xbold)}}$, 
\begin{align}
\beta_{\xbold} \ket{\phi_{\xbold}} & = \frac{1}{\sqrt{2^{|H|}}}  \sum_{\zbold} (-1)^{\underset{(k, l) \in E_H}{\oplus} z_k \land z_l} (-1)^{\underset{i}{\oplus} x_i \land z_i} \alpha_{\zbold} \ket{\psi_{\zbold}}, \text{   } \forall \xbold. 
\label{eq:beta}
\end{align}

Note that for certain $\xbold$, $\beta_{\xbold} = 0$. In Equation (\ref{eq:alpha}), some terms will then be zero. In Equation (\ref{eq:beta}), the whole left-hand side will then be zero. 
If any of the expressions are equal to zero, we can use them to show that the terms are grouped as in the Schmidt decomposition of $\ket{\mathscr{G}}$ in Equation (\ref{eq:schmitty}), which will allow us to simplify Equations  (\ref{eq:beta}) and (\ref{eq:compbas}). If none of the expressions are equal to zero, we will see that the same simplification applies. This is encapsulated in Lemma \ref{l:subl}.
\begin{manuallemma}{1.1A}
Equations (\ref{eq:beta}) and (\ref{eq:compbas}) can be written as
\begin{align}
\text{\ \ } \beta_{\vbold} \ket{\phi_{\vbold}} & = \frac{\sqrt{2^{|H|}}}{s} \sum_{\wbold} (-1)^{\underset{i}{\oplus} v_i \land w_i} \alpha_{\wbold} \ket{\psi_{\wbold}}, \text{ \ } \forall \vbold, \\
\ket{\Psi} & = \sum_{\zbold} (-1)^{\underset{(k, l) \in E_H}{\oplus} z_k \land z_l} \alpha_{\wbold} \ket{\zbold}_H \otimes \ket{\psi_{\wbold}}_D,
\end{align}
where $\wbold, \vbold \in \{ bn(0), ..., bn(s-1) \}$.
\label{l:subl}
\end{manuallemma}
\begin{proof}
We will prove this in a series of steps. First, we show that the $\ket{\psi_{\zbold}}$s are equal for the $\ket{\zbold}$ terms that are grouped together in the Schmidt decomposition of $\ket{\mathscr{G}}$ in Equation (\ref{eq:schmitty}), and that their corresponding $\alpha_{\zbold}$s are equal up to $\pm 1$. 
For any strings $\zbold, \zbold' \text{ such that } \forall d \in D, \underset{i \in N(d)}{\oplus} z_i = \underset{i \in N(d)}{\oplus} z'_i$, we have using Equation (\ref{eq:alpha}), 
\begin{align}
(-1)^{\underset{(k, l) \in E_H}{\oplus} z_k \land z_l} \alpha_{\zbold} \ket{\psi_{\zbold}} -  (-1)^{\underset{(k, l) \in E_H}{\oplus} z_k' \land z_l'} \alpha_{\zbold'} \ket{\psi_{\zbold'}} 
 = \frac{1}{\sqrt{2^{|H|}}} \sum_{\xbold} \beta_{\xbold} \Big[ (-1)^{\underset{i}{\oplus} x_i \land z_i} - (-1)^{\underset{i}{\oplus} x_i \land z_i'} \Big] \ket{\phi_{\xbold}}. 
\label{eq:apple}
\end{align}
From Equation (\ref{eq:bxz}), we know that for such strings $\zbold, \zbold'$, if $\underset{i}{\oplus} (z_i \oplus z_i' ) \land x_i = 1$, then $\beta_{\xbold} = 0$.
Otherwise, for the terms where $\beta_{\xbold} \neq 0$, we have
\begin{align}
0 & = \underset{i}{\oplus} (z_i \oplus z_i' ) \land x_i  = \underset{i}{\oplus} \Big[ (x_i \land z_i) \oplus (x_i \land z_i') \Big]  = \Big[ \underset{i}{\oplus} (x_i \land z_i) \Big] \oplus \Big[ \underset{i}{\oplus} (x_i \land z_i') \Big], 
\end{align}
which gives $\underset{i}{\oplus} x_i \land z_i  = \underset{i}{\oplus} x_i \land z_i'$, and subsequently 
\begin{align}
(-1)^{\underset{i}{\oplus} x_i \land z_i}  = (-1)^{\underset{i}{\oplus} x_i \land z_i'}.
\label{eq:minusx}
\end{align}
Substituting in Equation (\ref{eq:apple}), we get that
$\forall \zbold, \zbold' \text{ such that } \forall d \in D, \underset{i \in N(d)}{\oplus} z_i = \underset{i \in N(d)}{\oplus} z'_i$, 
\begin{align}
(-1)^{\underset{(k, l) \in E_H}{\oplus} z_k \land z_l} \alpha_{\zbold} \ket{\psi_{\zbold}} & = (-1)^{\underset{(k, l) \in E_H}{\oplus} z'_k \land z'_l} \alpha_{\zbold'} \ket{\psi_{\zbold'}}. 
\label{eq:sol}
\end{align}



%

\noindent Next, we will simplify the expression for $\beta_{\xbold} \ket{\phi_{\xbold}}$, by substituting
Equations (\ref{eq:sol}) and (\ref{eq:minusx}) in Equation (\ref{eq:beta}) to get
\begin{align}
\beta_{\xbold} \ket{\phi_{\xbold}} = \ & \frac{1}{\sqrt{2^{|H|}}}  \sum_{\zbold} (-1)^{\underset{(k, l) \in E_H}{\oplus} z_k \land z_l} (-1)^{\underset{i}{\oplus} x_i \land z_i} \alpha_{\zbold} \ket{\psi_{\zbold}}, \text{   } \forall \xbold \text{ such that } \beta_{\xbold} \neq 0  \nonumber \\
= \ & \frac{1}{\sqrt{2^{|H|}}} \Big[ \sum_{\substack{\zbold \text{ s. t. } \\ \underset{i \in N(d_1)}{\oplus} z_i = 0 \land ... \\ \land \underset{i \in N(d_{|D|})}{\oplus} z_i = 0}} (-1)^{\underset{(k, l) \in E_H}{\oplus} z_k \land z_l} (-1)^{\underset{i}{\oplus} x_i \land z_i} \alpha_{\zbold} \ket{\psi_{\zbold}} \nonumber \\
& + ... + \sum_{\substack{\zbold \text{ s. t. } \\ \underset{i \in N(d_1)}{\oplus} z_i = 1 \land ... \\ \land \underset{i \in N(d_{|D|})}{\oplus} z_i = 1}} (-1)^{\underset{(k, l) \in E_H}{\oplus} z_k \land z_l} (-1)^{\underset{i}{\oplus} x_i \land z_i} \alpha_{\zbold} \ket{\psi_{\zbold}} \Big].
\end{align}
There are $s$ non-zero expressions for $\beta_{\xbold} \ket{\phi_{\xbold}}$, since there are $s$ non-zero $\beta_{\xbold}$s. 
There are also $s$ sum terms on the right-hand side of the above expression.
There are a total of $2^{|H|}$ terms of $\alpha_{\zbold} \ket{\psi_{\zbold}}$, grouped together if $\forall d \in D, \underset{i \in N(d)} {\oplus} z_i$ is the same. This means that, within each of the $s$ sum terms, there are $\frac{2^{|H|}}{s}$ terms. 
From Equation (\ref{eq:sol}), we know that $(-1)^{\underset{(k, l) \in E_H}{\oplus} z_k \land z_l} \alpha_{\zbold} \ket{\psi_{\zbold}}$ is the same for each of these terms. 
Further, we know from Equation (\ref{eq:minusx}) that $(-1)^{\underset{i}{\oplus} x_i \land z_i}$ is the same for each of these terms (since $\beta_{\xbold} \neq 0$). Then, we can write our expression using $s$ terms (corresponding to each sum term in the above expression). 

In this way, we are left with terms depending on $s$ unique variables that can take any value 0 or 1, instead of $2^{|H|}$ variables that can only take certain values. 
Defining new variables $\wbold, \vbold \in \{ bn(0), ..., bn(s-1) \}$, we can write the above as 
\begin{align}
\beta_{\vbold} \ket{\phi_{\vbold}} =  \frac{1}{\sqrt{2^{|H|}}}  \frac{2^{|H|}}{s} \sum_{\wbold} (-1)^{\underset{i}{\oplus} v_i \land w_i} \alpha_{\wbold} \ket{\psi_{\wbold}} = \frac{\sqrt{2^{|H|}}}{s} \sum_{\wbold} (-1)^{\underset{i}{\oplus} v_i \land w_i} \alpha_{\wbold} \ket{\psi_{\wbold}}, \text{ \ \ } \forall \vbold. 
\label{eq:yak}
\end{align}
Note that $\forall \wbold \neq \wbold',$ $\bra{\psi_{\wbold}}\ket{\psi_{\wbold'}} = 0$, from Equation (\ref{eq:psiorth}) and the fact that any two states with different $\wbold$ have different $\underset{i \in N(d)}{\oplus} z_i$ for some $d \in D$. As before, $\forall \vbold \neq \vbold'$, $\bra{\phi_{\vbold}}\ket{\phi_{\vbold'}} = 0$, from Equation (\ref{eq:phiorth}). 
This means that now, instead of having $2^{|H|}$ variables $\zbold, \xbold \in \{ bn(0), ..., bn(2^{|H|} - 1) \}$, we have $s$ variables $\wbold, \vbold \in \{ bn(0), ..., bn(s-1) \}$.

Finally, we simplify the computational basis expression for $\ket{\Psi}$ in Equation (\ref{eq:compbas}). The terms should now be grouped as in Equation (\ref{eq:schmitty}), with $s$ terms, as
\begin{align}
\ket{\Psi}  = \sum_{\zbold} \alpha_{\zbold} \ket{\zbold}_H \otimes \ket{\psi_{\zbold}}_D  
 = \sum_{\zbold} (-1)^{\underset{(k, l) \in E_H}{\oplus} z_k \land z_l} \alpha_{\wbold} \ket{\zbold}_H \otimes \ket{\psi_{\wbold}}_D,
 \label{eq:orange}
\end{align}
where $\wbold = f(\zbold)$ depends on $\underset{i \in N(d)}{\oplus} z_i, \forall d \in D$. 

If none of the expressions are equal to zero, we have $s = 2^{|H|}$, and so we can simply rename $\zbold$ as $\wbold$, as they take values from the same set $\{ bn(0), ..., bn(s-1) \}$.
\end{proof}

Our next aim is to determine the value of $\alpha_{\wbold}$. From the normalisation condition, $\bra{\Psi}\ket{\Psi} = 1$, we have
\begin{align}
\sum_{\wbold} \frac{2^{|H|}}{s} \abs{\alpha_{\wbold}}^2 = 1 \implies 
\sum_{\wbold} \abs{\alpha_{\wbold}}^2  = \frac{s}{2^{|H|}},
\label{eq:normnorm}
\end{align}
since there are $\frac{2^{|H|}}{s}$ terms of $\ket{\zbold}$ corresponding to each $\wbold$, as we saw before.

To get the orthogonality conditions, we take the overlap of each non-zero expression for $\beta_{\vbold} \ket{\phi_{\vbold}}$ with another non-zero expression for $\beta_{\vbold'} \ket{\phi_{\vbold'}}$, to get $(s-1)$ equations for all $\vbold \neq \vbold'$ given by
\begin{align}
\beta_{\vbold}^\dag \beta_{\vbold'} \bra{\phi_{\vbold}}\ket{\phi_{\vbold'}} = 0 & = \frac{2^{|H|}}{s^2} \Big[ \sum_{\wbold}  (-1)^{\underset{i}{\oplus} v_i \land w_i} (-1)^{\underset{i}{\oplus} v'_i \land w_i} \abs{\alpha_{\wbold}}^2 \bra{\psi_{\wbold}}\ket{\psi_{\wbold}} \Big],
\end{align}
which we can write as
\begin{align}
 \sum_{\wbold} (-1)^{\underset{i}{\oplus} p_i \land w_i} \abs{\alpha_{\wbold}}^2 = 0, \text{ \ \ } \forall \pbold \in \{ bn(1), ..., bn(s-1) \}.
 \label{eq:orthorth}
\end{align}
(Note that since the overlap is taken with two different expressions, $\pbold$ can never be $0...0$.)

We will now solve the $s$ equations for $s$ variables (from the normalisation and orthogonality conditions) using the matrix method, formulating the system of equations as $\mathcal{A} u = b$. 
The matrix $\mathcal{A}$ is of size $s \times s$, while $u, b$ are of size $s \times 1$. $u$ is a column vector containing each $\abs{\alpha_{\wbold}}^2$, and $b$ is a column vector giving the normalisation and orthogonality conditions. 
$\mathcal{A}$ has 1s in its first row and column, and the other elements are $\pm 1$. The 1s on the first row are from the normalisation condition. The 1s on the first column occur because for $\wbold = 0...0$, the exponent of $(-1)$ will always be 0 (since the AND of anything with $0...0$ gives $0...0$), and so the sign of $\abs{\alpha_{\wbold}}^2$ is always +1. 
We then have
\begin{align}
\mathcal{A} = 
\begin{bmatrix}
1 & 1 & ... & 1 \\[0.3em] 
1 & \pm 1 & ... & \pm 1 \\[0.3em]
... & ... & ... & ... \\[0.3em]
1 & \pm 1 & ... & \pm 1 
\end{bmatrix},
u = 
\begin{bmatrix} 
\abs{\alpha_{00...00}}^2 \\[0.3em]
\abs{\alpha_{00...01}}^2 \\[0.3em]
... \\[0.3em] 
\abs{\alpha_{11...11}}^2 
\end{bmatrix}, 
b = \frac{s}{2^{|H|}}
\begin{bmatrix}
1 \\[0.3em]
0 \\[0.3em]
... \\[0.3em]
0 
\end{bmatrix}.
\end{align}
The values of $\abs{\alpha_{\wbold}}^2$ are then determined by 
\begin{align}
\begin{bmatrix} 
\abs{\alpha_{00...00}}^2 \\[0.3em]
\abs{\alpha_{00...01}}^2 \\[0.3em]
... \\[0.3em] 
\abs{\alpha_{11...11}}^2 
\end{bmatrix}  
= 
 \mathcal{A}^{-1} \frac{s}{2^{|H|}} \begin{bmatrix}
1 \\[0.3em]
0 \\[0.3em]
... \\[0.3em]
0 
\end{bmatrix}
= \frac{s}{2^{|H|}} \begin{bmatrix}
\mathcal{A}^{-1}_{1,1} \\[0.3em]
\mathcal{A}^{-1}_{2,1} \\[0.3em]
... \\[0.3em]
\mathcal{A}^{-1}_{s, 1} 
\end{bmatrix}.
\label{eq:matrixsol}
\end{align}
Thus, the solution to the set of equations is equal to $\frac{s}{2^{|H|}}$ times the first column of $\mathcal{A}^{-1}$.
Using Equations (\ref{eq:normnorm}) and (\ref{eq:orthorth}), let us now write more precisely what the elements of $\mathcal{A}$ are:
\begin{align}
\mathcal{A} & = 
\begin{bmatrix}
(-1)^{\oplus 00...00 \land 00...00} & (-1)^{\oplus 00...00 \land 00...01} & ... & (-1)^{\oplus 00...00 \land 11...11} \\[0.3em]
(-1)^{\oplus 00...01 \land 00...00} & (-1)^{\oplus 00...01 \land 00...01} & ... & (-1)^{\oplus 00...01 \land 11...11} \\[0.3em]
... & ... & ... & ... \\[0.3em]
(-1)^{\oplus 11...11 \land 00...00} & (-1)^{\oplus 11...11 \land 00...01} & ... & (-1)^{\oplus 11...11 \land 11...11} \\[0.3em]
\end{bmatrix},
\end{align}
where by $\oplus \mathsf{a} \land \mathsf{b}$ we mean computing the AND operation of the strings $\mathsf{a}, \mathsf{b}$, and then XORing all the resulting bits (or, in other words, finding the parity of the AND of the strings $\mathsf{a}, \mathsf{b}$).

Let us take $\pbold_i, \wbold_j$ to be the $i^{\text{th}}$ and $j^{\text{th}}$ strings in the set $\{ bn(0), ..., bn(s-1) \}$, where $i, j \in \{1, ..., s\}$. (For example, if $i=2, j=3$, we have $\pbold_2 = 00...01, \wbold_3 = 00...10$.) Then, taking $i$ to denote the row and $j$ to denote the column, $\mathcal{A}$ is given by
\begin{align}
\mathcal{A} 
&  = \begin{bmatrix}
(-1)^{\oplus \pbold_1 \land \wbold_1} & (-1)^{\oplus \pbold_1 \land \wbold_2} & ... & (-1)^{\oplus \pbold_1 \land \wbold_{s}} \\[0.3em]
(-1)^{\oplus \pbold_2 \land \wbold_1} & (-1)^{\oplus \pbold_2 \land \wbold_2} & ... & (-1)^{\oplus \pbold_2 \land \wbold_{s}} \\[0.3em]
... & ... & ... & ... \\[0.3em]
(-1)^{\oplus \pbold_{s} \land \wbold_1} & (-1)^{\oplus \pbold_{s} \land \wbold_2} & ... & (-1)^{\oplus \pbold_{s } \land \wbold_{s}} \\[0.3em]
\end{bmatrix}.
\end{align}
In general, the $(i,j)^{\text{th}}$ element of $\mathcal{A}$ is given by 
\begin{align}
\mathcal{A}_{i, j} = (-1)^{\oplus \pbold_i \land \wbold_j}.
\end{align}
Note that since $\pbold_i = \wbold_i \forall i$, we have $
\mathcal{A}_{i,j} = (-1)^{\oplus \pbold_i \land \wbold_j}  = (-1)^{\oplus \wbold_i \land \pbold_j}  = (-1)^{\oplus \pbold_j \land \wbold_i} = \mathcal{A}_{j, i}$,
and so $\mathcal{A}$ is symmetric.
We will now show that the inverse of the matrix $\mathcal{A}$ is given by $\mathcal{A}^{-1} = \frac{1}{s} \mathcal{A}$. To prove this, we just have to show that $\mathcal{A} \mathcal{A}^{-1} = \mathds{1}$. Denoting $\mathcal{A}\mathcal{A} = \mathcal{M}$, we have 
\begin{align}
\mathcal{A} \mathcal{A}^{-1}  = \mathcal{A} \frac{1}{s} \mathcal{A}  = \frac{1}{s} \mathcal{A}\mathcal{A} = \frac{1}{s} \mathcal{M}.
\label{eq:ainv}
\end{align} 
The $(i,j)^{\text{th}}$ element of $\mathcal{M}$ is given by
\begin{align}
\mathcal{M}_{i, j} & =  \sum_{k = 1}^{s} (-1)^{\oplus \pbold_i \land \wbold_k} (-1)^{\oplus \pbold_k \land \wbold_j}  = \sum_{k=1}^{s} (-1)^{\oplus \pbold_i \land \wbold_k} (-1)^{\oplus  \pbold_j \land \wbold_k}   = \sum_{k=1}^{s} (-1)^{\oplus (\pbold_i \oplus  \pbold_j ) \land \wbold_k}.
\end{align}
When $i = j$, we have $\pbold_i \oplus \pbold_j = 00...00$, and so for every $\wbold_k$, we always have $(-1)^0$. Thus, we have $\mathcal{M}_{i, i} = \sum_{k=1}^{s} 1 = s$.
When $i \neq j$, the AND operation is performed on the same string $\pbold_i \oplus \pbold_j$ and $\wbold_k, \forall k$. In this case, $\pbold_i \oplus \pbold_j \in \{ bn(1), ..., bn(s-1) \}$, as their sum can only be $00...00$ if $i = j$. Thus, if we sum over the AND of this string with each possible $\wbold \in \{ 00...00, ..., 11...11 \}$, we get an equal number of $+1$s and $-1$s. So, this sum is equal to 0. We can therefore say
\begin{align}
\mathcal{M}_{i, j} = \delta_i^j s,
\end{align}
which means $\mathcal{M} = s \mathds{1}$. Substituting in Equation (\ref{eq:ainv}), we find $\mathcal{A} \mathcal{A}^{-1} = \mathds{1}$.
Thus, $\mathcal{A}^{-1} =  \frac{1}{s} \mathcal{A}$, which we substitute in Equation (\ref{eq:matrixsol}) to get 
\begin{align}
\begin{bmatrix} 
\abs{\alpha_{0...0}}^2 \\[0.3em]
\abs{\alpha_{0...1}}^2 \\[0.3em]
... \\[0.3em] 
\abs{\alpha_{1...1}}^2 
\end{bmatrix}  
= \frac{s}{2^{|H|}} \begin{bmatrix}
\mathcal{A}^{-1}_{1,1} \\[0.3em]
\mathcal{A}^{-1}_{2,1} \\[0.3em]
... \\[0.3em]
\mathcal{A}^{-1}_{s, 1} 
\end{bmatrix} 
=  \frac{1}{s} \times \frac{s}{2^{|H|}} \begin{bmatrix}
\mathcal{A}_{1,1} \\[0.3em] \mathcal{A}_{2,1} \\[0.3em] ... \\[0.3em] \mathcal{A}_{s, 1} \\[0.3em]
\end{bmatrix}
= \frac{1}{s} \times \frac{s}{2^{|H|}} \begin{bmatrix}
1 \\[0.3em] 1 \\[0.3em] ... \\[0.3em] 1 \\[0.3em]
\end{bmatrix}
=  \begin{bmatrix}
\frac{1}{2^{|H|}} \\[0.3em] \frac{1}{2^{|H|}} \\[0.3em] ... \\[0.3em] \frac{1}{2^{|H|}} \\[0.3em]
\end{bmatrix}.
\end{align}
This gives each $\alpha_{\wbold} = \pm \frac{1}{\sqrt{2^{|H|}}}$.
Let us substitute the value of $\alpha_{\wbold}$ in Equation (\ref{eq:orange}). We get
\begin{align}
\ket{\Psi} & = \pm \frac{1}{\sqrt{2^{|H|}}} \sum_{\zbold} (-1)^{\underset{(k, l) \in E_H}{\oplus} z_k \land z_l} \ket{\zbold}_H \otimes \ket{\psi_{\wbold}}_D  \nonumber \\
& =  \pm \frac{1}{\sqrt{2^{|H|}}} \sum_{\zbold} (-1)^{\underset{(k, l) \in E_H}{\oplus} z_k \land z_l} \ket{\zbold}_H \otimes \ket{\psi_{(f(\zbold))}}_D.
\end{align}

Finally, $\rho_H^{\ket{\Psi}}$ only depends on the honest Schmidt basis. Comparing the above expression with Equation (\ref{eq:schmitty}) shows that the honest Schmidt basis of the actual state $\ket{\Psi}$ is equal to the honest Schmidt basis of the ideal state $\ket{\mathscr{G}}$, giving $\rho_H^{\ket{\Psi}} = \rho_H^{\ket{\mathscr{G}}}$. If $\ket{\Psi}$ is a purification of $\rho_H^{\ket{\Psi}}$, and $\ket{\mathscr{G}}$ is a purification of $\rho_H^{\ket{\mathscr{G}}}$, then due to the unitary equivalence of purifications, we must have $\ket{\Psi} = U_D \ket{\mathscr{G}}$ if all the test measurements pass perfectly.
This concludes the proof of Lemma \ref{lem:windsurf}.
\end{proof}

The above Lemma \ref{lem:windsurf} tells us that only the state $U_D \ket{\mathscr{G}}$ (the ideal state up to local unitaries on the dishonest side), or equivalently $\rho_H^{\ket{\mathscr{G}}}$, always passes the stabiliser test $\mathcal{S}_j$ in any round $j$. 
For simplicity of presentation later on, we now analyse the case where the number of stabilisers that need to be measured to have the analogous statement of Lemma \ref{lem:windsurf} is $\mathsf{J}$.
We fix $N_{total}$ such that $\mathsf{N} \equiv N_{total} - \mathsf{J} N_{test} = \mathsf{J} N_{test}$ is the total number of remaining copies, out of which one is chosen 
to be the target copy.
Let $\mathsf{k}$ be the number of copies out of the remaining $\mathsf{N}$ copies that are $\rho_H^{\ket{\mathscr{G}}}$.
Now, using the Serfling bound,
we will find a bound on $\frac{\mathsf{k}}{\mathsf{N}}$
in the following Lemma, which is an adaptation of the method of Takeuchi et al. \cite{Takeuchi2019} to our scenario of verifying general graph states with any number of dishonest parties.
\begin{manuallemma}{1.2}[adapted from {\cite{Takeuchi2019}}]
The probability that the fraction of states that would pass all stabiliser tests out of the remaining copies in Protocol \ref{alg:takeuchi} is given by 
\begin{align}
\frac{\mathsf{k}}{\mathsf{N}} 
\geq 1 - \frac{2\sqrt{c}}{\mathsf{J}} - 2 \mathsf{J} \Big( 1 - \frac{N_{pass}}{\mathsf{J} N_{test}} \Big)
\end{align}
is at least $1 - \mathsf{J}^{1-\frac{2cm}{3}}$.
\label{lem:fullstablem}
\end{manuallemma}
\begin{proof}

Consider a set of binary random variables $Y = \{ Y_1, ..., Y_{\mathcal{T}} \}$ where $Y_t \in \{ 0, 1 \}$. Let us set the value of $Y_t = 0$ if the stabiliser test passes on copy $t$, and otherwise $Y_t = 1$. Then, by Serfling's bound \cite{Serfling1974}, where $\mathcal{T}= \mathcal{L}+\mathcal{R}$, for any $0 < \nu < 1$, 
\begin{align}
\text{Pr} \Big[ \sum_{t \in \overline{\mathsf{\Pi}} } Y_t \leq \frac{\mathcal{L}}{\mathcal{R}} \sum_{t \in \mathsf{\Pi}} Y_t + \mathcal{L}\nu \Big] 
\geq 1 - \exp \big[ - \frac{2 \nu^2 \mathcal{L} \mathcal{R}^2}{(\mathcal{L}+\mathcal{R})(\mathcal{R}+1)} \big],
\label{eq:exp}
\end{align}
where $\mathsf{\Pi}$ is a set of $\mathcal{R}$ samples chosen independently and uniformly randomly from $Y$ without replacement, and $\overline{\mathsf{\Pi}}$ is the complementary set of $\mathsf{\Pi}$.
The expression inside the probability bracket is then an upper bound on the number of copies out of the remaining that would fail the stabiliser test, given a number of copies that pass. 

Now, let us consider the stabiliser measurement $\mathcal{S}_{j \in \{1, ..., \mathsf{J}\} }$. We take $\mathcal{L} = N_{total} - jN_{test}$, and $\mathcal{R} = N_{test}$. Let us then set $\nu = \frac{\sqrt{c}}{\mathsf{J}^2}$, which is chosen in this way to maximise both the fidelity and probability in our resulting expression (this implies $0 < \sqrt{c} < \mathsf{J}^2$). Let $\mathsf{\Pi}^{(j)}$ be the set of copies on which each $\mathcal{S}_j$ was measured, and $\overline{\mathsf{\Pi}}^{(j)}$ be the set of remaining copies after measuring $\mathcal{S}_j$. Finally, we denote the probability expression on the right-hand side of Equation (\ref{eq:exp}) for the stabiliser test in round $j$ as $q_j$. Then, after the stabiliser test for $\mathcal{S}_j$ is performed, we have 
\begin{align} \label{eq q_j}
\text{Pr} \Big[ \sum_{t \in \overline{\mathsf{\Pi}}^{(j)} } Y_t \leq \frac{N_{total} - j N_{test}}{N_{test}} \sum_{t \in \mathsf{\Pi}^{(j)}} Y_t + (N_{total} - j N_{test}) \nu \Big]
\geq q_j.
\end{align}
In other words, with probability at least $q_j$, the maximum number of copies that would fail the stabiliser test of $\mathcal{S}_j$ out of the remaining is given by $\frac{N_{total} - j N_{test}}{N_{test}} \sum_{t \in \mathsf{\Pi}^{(j)}} Y_t+ (N_{total} - j N_{test})\nu$. Then, the minimum number of copies that would pass the $\mathcal{S}_j$ test is given by the total number of remaining copies minus this maximum number of failed copies. 
Thus, after all stabiliser tests are performed, we can lower bound the number of remaining copies which would pass all of the stabiliser tests by summing over all $j$'s corresponding to all stabilisers $\mathcal{S}_j$. 
Recalling that we denote the number of `good' remaining copies by $\mathsf{k}$, we then have
\begin{align}
\mathsf{k} & \geq (N_{total} - \mathsf{J} N_{test}) - \sum_{j=1}^{\mathsf{J}} \Bigg[ \frac{N_{total} - j N_{test}}{N_{test}}  \sum_{t \in \mathsf{\Pi}^{(j)}} Y_t + (N_{total} - j N_{test}) \nu \Bigg]  \nonumber \\
& = 2\mathsf{J} N_{test} - \mathsf{J} N_{test} - \frac{N_{total}}{N_{test}} \sum_{j=1}^{\mathsf{J}} \sum_{t \in \mathsf{\Pi}^{(j)}} Y_t + \sum_{j=1}^{\mathsf{J}} j \sum_{t \in \mathsf{\Pi}^{(j)}} Y_t - \nu \mathsf{J} N_{total} + \sum_{j=1}^{\mathsf{J}} j N_{test} \nu  \nonumber \\
& \geq \mathsf{J} N_{test} - \nu \mathsf{J} N_{total} - \frac{N_{total}}{N_{test}} \sum_{j=1}^{\mathsf{J}} \sum_{t \in \mathsf{\Pi}^{(j)}} Y_t  \nonumber \\
& \geq \mathsf{J} N_{test} - \nu \mathsf{J} N_{total} - \frac{N_{total}}{N_{test}} (\mathsf{J} N_{test} - N_{pass})  \nonumber \\
& = \mathsf{J} N_{test} - \nu \mathsf{J} \times 2 \times \mathsf{J} N_{test} - \frac{2\mathsf{J} N_{test}}{N_{test}} (\mathsf{J} N_{test} - N_{pass})  \nonumber \\
& = \Big( \mathsf{J} - 2\sqrt{c} - 2 \mathsf{J}^2 + 2 \mathsf{J} \frac{N_{pass}}{N_{test}} \Big) N_{test},
\end{align}
which gives the fraction of `good' remaining copies, $\frac{\mathsf{k}}{\mathsf{N}}$, as
\begin{align}
\frac{\mathsf{k}}{\mathsf{N}} 
& \geq \frac{ \Big( \mathsf{J} - 2\sqrt{c} - 2\mathsf{J}^2 + 2 \mathsf{J} \frac{N_{pass}}{N_{test}} \Big)  N_{test} }{N_{total} - \mathsf{J}N_{test}}  \nonumber \\
& = \frac{\mathsf{J} - 2\sqrt{c} - 2\mathsf{J}^2 + 2\mathsf{J} \frac{N_{pass}}{N_{test}} }{\mathsf{J}}  \nonumber \\
& = 1 - \frac{2\sqrt{c}}{\mathsf{J}} - 2 \mathsf{J} \Big( 1 - \frac{N_{pass}}{\mathsf{J} N_{test}} \Big).
\end{align}

From Equation (\ref{eq:exp}), we then get for $q_j$, the probability corresponding to the test $\mathcal{S}_j$, 
\begin{align} \label{eq qj bound}
q_j & =  1 - \exp \Bigg[  - \frac{2 \nu^2 (N_{total} - j N_{test}) N_{test}^2 } {(N_{total} - j N_{test} + N_{test}) (N_{test} + 1) } \Bigg]  \nonumber \\
& =  1 - \exp \Bigg[ - 2 \nu^2 N_{test} \frac{1}{\frac{N_{total} - j N_{test} + N_{test}}{N_{total} - j N_{test}} \times \frac{N_{test} + 1}{N_{test}}} \Bigg]  \nonumber \\
& = 1 - \exp \Bigg[ - 2 \nu^2 N_{test} \frac{1}{1 + \frac{1}{2 \mathsf{J} -j}} \frac{1}{1 + \frac{1}{N_{test}}} \Bigg].
\end{align}
Now, let us compute the total probability corresponding to the full set of $\mathsf{J}$ stabiliser tests.
Taking $N_{test} = m \mathsf{J}^4 \ln \mathsf{J}$ and using $\mathsf{J} \geq 2
\implies \frac{1}{1 + \frac{1}{\mathsf{J}}} \geq \frac{2}{3}$, and 
$N_{test} \geq 1 
\implies \frac{1}{1 + \frac{1}{N_{test}}} \geq \frac{1}{2}$, we get the total probability corresponding to $\mathsf{J}$ stabiliser tests as
\begin{align}
\prod_{j=1}^{\mathsf{J}} q_j  \geq q_{\mathsf{J}}^{\mathsf{J}}  
& = \Bigg[ 1 - \exp \Big( - 2 \nu^2 N_{test} \frac{1}{1 + \frac{1}{2 \mathsf{J} - \mathsf{J}}} \frac{1}{1 + \frac{1}{N_{test}}} \Big) \Bigg]^{\mathsf{J}}  \nonumber \\
& \geq \Bigg[ 1 - \exp \Big( - \frac{2 \nu^2 N_{test}}{3}  \Big) \Bigg]^{\mathsf{J}}  \nonumber \\
& \geq \Bigg[ 1 - \exp \Big( - \frac{2\nu^2 m \mathsf{J}^4}{3} \ln{\mathsf{J}} \Big) \Bigg]^{\mathsf{J}}  \nonumber \\
& = \Bigg[ 1 - \mathsf{J}^{-\frac{2c m}{3}} \Bigg]^{\mathsf{J}} \nonumber \\
& \geq 1 - \mathsf{J}^{1-\frac{2c m}{3}}.
\end{align}
Thus, the event that 
the fraction of `good' copies that pass all the stabiliser tests $\mathcal{S}_j$ is lower bounded by 
$\frac{\mathsf{k}}{\mathsf{N}} \geq 1 - \frac{2\sqrt{c}}{\mathsf{J}} - 2\mathsf{J} \big( 1 - \frac{N_{pass}}{ \mathsf{J} N_{test}} \big)$ occurs with probability greater than or equal to $1 - \mathsf{J}^{1-\frac{2cm}{3}}$. 

\end{proof}

In the final step of the protocol, one copy is chosen, out of the remaining $\mathsf{N} \equiv N_{total} - \mathsf{J} N_{test}$ copies, 
to be the target copy.
Let us now finish the proof by proving Lemma \ref{lem:finalbit}, relating the above analysis to the averaged state of this 
target copy.

\begin{manuallemma}{1.3}
The probability that the fidelity of the averaged state of the 
target copy
in Protocol \ref{alg:takeuchi} satisfies
\begin{align}
F(\rho_H^{\ket{\mathscr{G}}}, \rho_H^{avg}) \geq \frac{\mathsf{k}}{\mathsf{N}}
\end{align}
is at least $ 1 - \mathsf{J}^{1-\frac{2cm}{3}}$.
\label{lem:finalbit}
\end{manuallemma}
\begin{proof}
Now, $\rho_H^{avg}$ is the reduced state of the honest parties of the averaged state, where the average is taken over all uniformly random selections of the 
target copy 
from the total number of remaining copies $\mathsf{N}$, 
and given by
\begin{align}
\rho_H^{avg} =  \frac{1}{\mathsf{N}} \sum_{i=1}^{\mathsf{N}} \rho_H^i.
\end{align}
Here, $\rho_H^i$ denotes the reduced state of the honest parties in each round (for each one of the remaining copies).
We now want to find the fidelity $F(\rho_H^{\ket{\mathscr{G}}}, \rho_H^{avg})$ between this averaged state and the ideal state (in terms of the honest reduced states). 
From the concavity of fidelity,
\begin{align}
F(\rho, \sum_i p_i \sigma_i) \geq \sum_i p_i F(\rho, \sigma_i).
\end{align}
We can use this to get
\begin{align}
F(\rho_H^{\ket{\mathscr{G}}}, \rho_H^{avg}) = F(\rho_H^{\ket{\mathscr{G}}}, \frac{1}{\mathsf{N}} \sum_{i=1}^{\mathsf{N}}  \rho_H^i)  \geq \frac{1}{\mathsf{N}} \sum_{i=1}^{\mathsf{N}} F(\rho_H^{\ket{\mathscr{G}}}, \rho_H^i).
\end{align}
Now, we know that for a number $\mathsf{k}
$ of the remaining states, we have $\rho_H^i = \rho_H^{\ket{\mathscr{G}}}$ with probability at least $ 1 - \mathsf{J}^{1-\frac{2cm}{3}}$. Assuming the worst case scenario that the rest of the states, each denoted by $\rho_H^{unknown}$, have zero fidelity with the ideal state, we get
\begin{align}
F(\rho_H^{\ket{\mathscr{G}}}, \rho_H^{avg}) & \geq \frac{1}{\mathsf{N}} \Big[ \mathsf{k} F(\rho_H^{\ket{\mathscr{G}}},  \rho_H^{\ket{\mathscr{G}}}) + (\mathsf{N}-\mathsf{k}) F(\rho_H^{\ket{\mathscr{G}}}, \rho_H^{unknown}) \Big]   \nonumber \\
& \geq \frac{1}{\mathsf{N}} \Big[ \mathsf{k} \times 1 + (\mathsf{N}-\mathsf{k}) \times 0 \Big]   \nonumber \\
& = \frac{\mathsf{k}}{\mathsf{N}}.
\end{align}
\end{proof}

Our final result then tells us that, with the appropriate choice of $N_{total}, N_{test}$, the parties can ensure that with probability at least $1 - \mathsf{J}^{1-\frac{2cm}{3}}$, 
the fidelity of the averaged state of the 
target copy
satisfies 
\begin{align}
F(\rho_H^{\ket{\mathscr{G}}}, \rho_H^{avg}) 
\geq 1 - \frac{2\sqrt{c}}{\mathsf{J}} - 2 \mathsf{J} \big( 1 - \frac{N_{pass}}{\mathsf{J} N_{test}} \big).
\end{align}
Note that in order to ensure that the probability and fidelity are greater than zero, the choice of the constants $m, c$ must be such that $\frac{3}{2m} < c < \frac{(\mathsf{J} - 1)^2}{4}$.
Setting $\mathsf{J}=2^n$, this concludes the proof of Theorem \ref{th:graphbigproof}.

\end{proof}

%
%
%
%

\newpage

\section{APPENDIX II: PROOF OF THEOREM \ref{th:allexamples}}

We now give some examples of types of graph states that are useful for specific purposes, and prove the following Theorem to show how we can reduce the resources required to verify them (in the presence of dishonest parties) in each case. 
\begin{manualtheorem}{2}
If $\ket{\mathscr{G}}$ is either
\begin{enumerate}
\item[A.] a complete graph state with dishonest parties anywhere in the network, or
\item[B.] a pentagon graph state with either one, three or four dishonest parties anywhere in the network, or two dishonest parties who are adjacent, or 
\item[C.] a cycle graph state with either one, $n-2$ or $n-1$ dishonest parties anywhere in the network, or any other number of dishonest parties who are adjacent, or 
\item[D.] a $\mathsf{1D}$ cluster state with either one or $n-1$ dishonest parties anywhere in the network, or any other number of adjacent honest and dishonest parties, or 
\item[E.] a $\mathsf{2D}$ cluster state with either one or $n-1$ dishonest parties anywhere in the network, or any other set of adjacent dishonest parties that forms a square or rectangle anywhere in the network,
\end{enumerate}
and we set $N_{total} = 2n N_{test}, N_{test} = \lceil{ m n^4 \ln n}\rceil$, 
and $\mathcal{S}$ as the set of stabiliser generators, assuming an honest Verifier,
the probability that the fidelity of the averaged state of 
the target copy (over all possible choices of the tested copies and target copy)
in Protocol \ref{alg:takeuchi} satisfies
\begin{align}
F(\rho_H^{\ket{\mathscr{G}}}, \rho_H^{avg}) 
\geq 1 - \frac{2\sqrt{c}}{n} - 2n \Big( 1 - \frac{N_{pass}}{n N_{test}} \Big) 
\end{align}
is at least $1 - n^{1 - \frac{2cm}{3}}$, where 
$m, c$ are positive constants chosen such that  $\frac{3}{2m} < c < \frac{(n-1)^2}{4}$.
\end{manualtheorem}

For clarity, we will treat each example separately. To prove such a statement, we start by inspecting the proof of Theorem \ref{th:graphbigproof}, to understand how the information about passing the stabiliser tests is used. In Lemma \ref{lem:windsurf}, this is firstly used to identify which elements of $\{ \ket{\psi_{\zbold}} \}, \{ \ket{\phi_{\xbold}} \}$ are orthogonal; however, as we see in Equations (\ref{eq:psiorth}) and (\ref{eq:phiorth}), this follows from passing purely the set of stabiliser generator measurements. It then remains to show that the conditions on which $\beta_{\xbold}$s are zero, given in Equation (\ref{eq:condition}), may also be deduced from only passing the stabiliser generator tests. (We point out that if the set $A$ is empty, which can be checked by examining the graph for the partition $(H, D)$, then from passing the full set of stabiliser measurements, we cannot set any $\beta_{\xbold}$ to be zero, and so we can trivially see that only the stabiliser generator measurements are required.) If this is true, it is possible for the parties to protect themselves against dishonest action even by running Protocol \ref{alg:takeuchi} with this simpler set of test measurements. 

Recall that once we know which $\beta_{\xbold}$s must be zero, we can determine which terms have the same $\alpha_{\zbold}, \ket{\psi_{\zbold}}$, allowing us to group the $\ket{\zbold}$ terms in Equation (\ref{eq:compbas}) in the same way as in the Schmidt decomposition of the ideal graph state, leading to the conclusion that $\ket{\Psi} = U_D \ket{\mathscr{G}}$.
Thus, once we have derived the conditions for $\beta_{\xbold}$ to be zero, we can simply continue with the remainder of the proof of Lemma \ref{lem:windsurf}. Inserting $\mathsf{J}=n$ into  Lemmas \ref{lem:fullstablem} and \ref{lem:finalbit}, we obtain the desired results.

In the following, we adapt our general proof in this way to cater to the specificities of complete graphs, cycle graphs and cluster states, with particular characteristics of the dishonest set of parties. Armed with this information, we will show that our parties can verify each graph state in an efficient way.

\subsection{Complete graph states}

Complete graph states, where every vertex is connected to every other vertex, are locally equivalent to GHZ states. As we have seen, such states are central to schemes for quantum anonymous transmission \cite{Christandl2005, Unnikrishnan2019}, as well as secret sharing \cite{Hillery1999}, metrology \cite{Toth2014}, and many other applications. A verification test for GHZ states was already proposed and analysed in \cite{Pappa2012}; however, we now approach this goal using the Serfling bound method. This subsection also serves to provide a comprehensive example of how our protocol and analysis work for this particular graph state, and so we will go through all steps of the proof.

\begin{figure*}[t]
\centering
\includegraphics[trim = 0cm 7cm 0cm 6.5cm, width=0.7\textwidth]{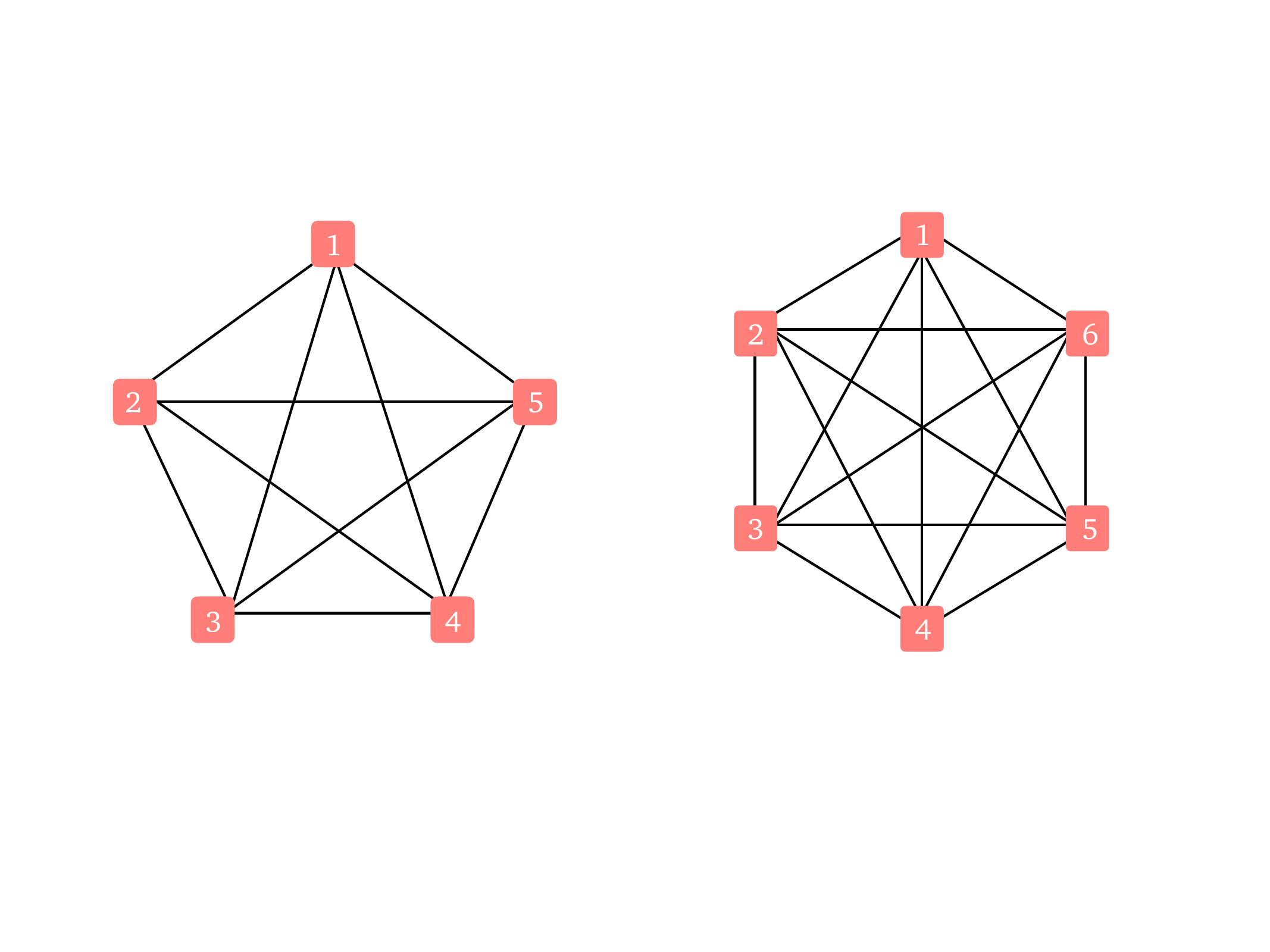}
\caption{5-qubit and 6-qubit complete graph states.}
\label{fig:complete}
\end{figure*}

Due to the symmetry of complete graphs (Figure \ref{fig:complete}), we will see that we can protect against any number of dishonest parties anywhere in the network by only measuring the stabiliser generators. Recall that we denote the stabiliser generator corresponding to the $i^{\text{th}}$ qubit as $K_i$. Let the parties now run Protocol \ref{alg:takeuchi} with $\mathcal{S}$ as the set of generators; our result is given in the following Theorem.

\begin{manualtheorem}{2A}


If $\ket{\mathscr{G}}$ is a complete graph state, and we set $N_{total} = 2n N_{test}$, $N_{test} = \lceil{ m n^4 \ln n}\rceil$, and $\mathcal{S}$ as the set of stabiliser generators, the probability that the fidelity of the averaged state of 
the target copy (over all possible choices of the tested copies and target copy) 
in Protocol \ref{alg:takeuchi} satisfies
\begin{align}
F(\rho_H^{\ket{\mathscr{G}}}, \rho_H^{avg}) 
\geq 1 - \frac{2\sqrt{c}}{n} - 2n \Big( 1 - \frac{N_{pass}}{n N_{test}} \Big) 
\end{align}
is at least $1 - n^{1 - \frac{2cm}{3}}$,
where 
$m, c$ are positive constants chosen such that $\frac{3}{2m} < c < \frac{(n-1)^2}{4}$.
\label{th:graphgenproof}
\end{manualtheorem}

\begin{proof}
As before, we prove this in stages, now for the specific case of complete graph states.
\begin{manuallemma}{2A.1}
The only state that can pass all stabiliser generator tests perfectly in each round is $\ket{\Psi} = U_D \ket{\mathscr{G}}$, where $U_D$ is a unitary on the dishonest part of the state. 
\end{manuallemma}
\begin{proof}
Complete graph states are of Schmidt rank 2, \emph{i.e.} there are always exactly two terms in their Schmidt decomposition. 
The Schmidt decomposition of a complete graph state with respect to any partition $(H, D)$ is 
\begin{align}
\ket{\mathscr{G}} = \ & \frac{1}{\sqrt{2^{|H|}}} \sum_{\zbold} (-1)^{\underset{(k, l) \in E_H}{\oplus} z_k \land z_l} \ket{\zbold}_H \otimes \prod_{a \in H} \big( \prod_{b \in N(a)} \sigma_Z^{(b)} \big)^{z_a} \ket{\mathscr{G-H}}_D \nonumber \\
= \ & \frac{1}{\sqrt{2^{|H|}}}  \Bigg[ \Big( \sum_{\Delta(\zbold) = 0 (\text{mod } 4)} \ket{\zbold} - \sum_{\Delta(\zbold) = 2 (\text{mod } 4)} \ket{\zbold} \Big)_H \otimes \ket{\mathscr{G-H}_{(00...00)}}_D \nonumber \\
& + \Big( \sum_{\Delta(\zbold) = 1 (\text{mod } 4)} \ket{\zbold} - \sum_{\Delta(\zbold) = 3 (\text{mod } 4)} \ket{\zbold} \Big)_H \otimes \ket{\mathscr{G-H}_{(11...11)}}_D \Bigg],
\end{align}
where $\ket{\mathscr{G-H}_{(00...00)}}$ is the dishonest subgraph with no $\sigma_Z$s on any vertices, and $\ket{\mathscr{G-H}_{(11...11)}}$ is the dishonest subgraph with $\sigma_Z$s on all vertices.
Thus, the reduced state of the honest parties of this ideal state is given by
\begin{align}
\rho_H^{\ket{\mathscr{G}}} = \ & \frac{1}{2^{|H|}} \Bigg[  \Big( \sum_{\Delta(\zbold) = 0 (\text{mod } 4)} \ket{\zbold} - \sum_{\Delta(\zbold) = 2 (\text{mod } 4)} \ket{\zbold} \Big)  \Big( \sum_{\Delta(\zbold) = 0 (\text{mod } 4)} \bra{\zbold} - \sum_{\Delta(\zbold) = 2 (\text{mod } 4)} \bra{\zbold} \Big) \nonumber \\
& + \Big( \sum_{\Delta(\zbold) = 1 (\text{mod } 4)} \ket{\zbold} - \sum_{\Delta(\zbold) = 3 (\text{mod } 4)} \ket{\zbold} \Big) \Big( \sum_{\Delta(\zbold) = 1 (\text{mod } 4)} \bra{\zbold} - \sum_{\Delta(\zbold) = 3 (\text{mod } 4)} \bra{\zbold} \Big) \Bigg].
\end{align}
From Equations (\ref{eq:compbas}) and (\ref{eq:subbas}), we know that our state can be written in general as $
\ket{\Psi}  = \sum_{\zbold} \alpha_{\zbold} \ket{\zbold}_H \otimes \ket{\psi_{\zbold}}_D$ and $
\ket{\Psi}  = \sum_{\xbold} \beta_{\xbold} \ket{\mathscr{H}_{(\xbold)}}_H \otimes \ket{\phi_{\xbold}}_D$. 
Let us now write the test measurements, which are the stabiliser generators here. Since, for the complete graph, each vertex shares an edge with every other vertex, the stabiliser generators are given by 
\begin{align}
K_i = X_i \prod_{e \in V} Z_e,
\end{align}
where $V$ is the set of vertices in the graph. 
We will again group the test measurements into Group 1, where the honest measurement only consists of $Z$s, and Group 2, which contains the remaining measurements. In order to pass perfectly, the overall outcome of each measurement must be $+1$.

Let us first consider the Group 1 measurements. Any $\ket{\zbold}$ that has an even number of 1s (parity of $\zbold$ is 0) will give outcome $+1$ for the honest measurement $Z...Z$, and so the dishonest measurement must give outcome $+1$ in order to pass the test. Any $\ket{\zbold}$ that has an odd number of 1s (parity of $\zbold$ is 1) will give outcome $-1$ for $Z...Z$, and so the dishonest measurement must give outcome $-1$. So, in order to pass perfectly, the dishonest parties must be able to discriminate perfectly between the states $\ket{\psi_{\Delta(\zbold) = 0 (\text{mod } 2)}}$ and $\ket{\psi_{\Delta(\zbold) = 1 (\text{mod } 2)}}$, where $\Delta(\zbold)$ is the Hamming weight of $\zbold$ (number of 1s in the string). This means that for all $\zbold$, we have $\bra{\psi_{\Delta(\zbold) = 0(\text{mod } 2)}}\ket{\psi_{\Delta(\zbold) = 1(\text{mod } 2)}} = 0$. 
(Note that this can be seen from the general proof, which tells us in this case that $\forall \zbold, \zbold'$ such that $\underset{i}{\oplus} z_i \neq \underset{i}{\oplus} z_i'$, we have $\bra{\psi_{\zbold}}\ket{\psi_{\zbold'}} = 0$.)

Now, let us see what happens if the Group 2 measurements pass perfectly. This group will have all measurements $XZ...Z, ZXZ...Z, ..., Z...ZX$ in the honest part, and the dishonest part of each will be $(Z...Z)'$ (so the same dishonest part for each). To pass perfectly, the dishonest parties must be able to always give the correct outcome; however, they only know that the measurement they must do is $(Z...Z)'$, and this does not tell them whether the honest parties are measuring $XZ...Z, ZXZ...Z, ..., Z...ZX$. In order for the dishonest parties to always output the correct outcome for all of these measurements (to make the overall outcome $+1$), the only terms that can appear in Equation (\ref{eq:subbas}) 
are the honest subgraph ($\ket{\mathscr{H}_{(00...00)}}$), and the honest subgraph with $\sigma_Z$s on all the vertices ($\ket{\mathscr{H}_{(11...11)}}$). So, passing the Group 2 measurements tells us that $\beta_{\xbold} = 0$ for $\xbold \notin \{ 00...00, 11...11 \}$. (Note that this is the same conclusion that can be derived from measuring the full set of stabilisers in the general proof. 
Every pair of honest vertices forms a possible set $A$ here, and we have $\beta_{\xbold} = 0$ if $\exists A$ such that $\underset{i \in A}{\oplus} x_i = 1$. 
The only remaining strings are then those $\xbold$ that have all elements equal to one another.)

We can then write the state in Equation (\ref{eq:subbas}) as
\begin{align}
\ket{\Psi} & = \beta_{00...00} \ket{\mathscr{H}_{(00...00)}}_H \ket{\phi_{00...00}}_D + \beta_{11...11} \ket{\mathscr{H}_{(11...11)}}_H \ket{\phi_{11...11}}_D.
\end{align}
Since measuring $(Z...Z)'$ on $\ket{\phi_{00...00}}$ must give outcome $+1$, and on $\ket{\phi_{11...11}}$ it must give outcome $-1$ (in order to always get the total outcome to be $+1$), the dishonest parties must be able to perfectly discriminate between these two states, and so we must have $\bra{\phi_{00...00}}\ket{\phi_{11...11}} = 0$.
Then, following the steps in the general proof, we get 
\begin{align}
\beta_{\xbold} \ket{\phi_{\xbold}} & = \frac{1}{\sqrt{2^{|H|}}} \Big[ \sum_{\zbold} (-1)^{\underset{(k, l) \in E_H}{\oplus} z_k \land z_l}  (-1)^{\underset{i}{\oplus} x_i \land z_i} \alpha_{\zbold} \ket{\psi_{\zbold}}_D \Big],  \text{      } \forall  \xbold \text{ such that } \beta_{\xbold} \neq 0, \\
0 & = \frac{1}{\sqrt{2^{|H|}}} \Big[ \sum_{\zbold} (-1)^{\underset{(k, l) \in E_H}{\oplus} z_k \land z_l} (-1)^{\underset{i}{\oplus} x_i \land z_i} \alpha_{\zbold} \ket{\psi_{\zbold}}_D \Big], \text{      } \forall  \xbold \text{ such that } \beta_{\xbold} = 0.
\label{eq:turtle}
\end{align}
The solution of Equation (\ref{eq:turtle}), as adapted for complete graphs from the general proof, is 
$\forall \zbold, \zbold'$ such that $ \Delta(\zbold) \text{ mod 2} = \Delta(\zbold') \text{ mod 2}$, we have
\begin{align}
(-1)^{\underset{(k, l) \in E_H}{\oplus} z_k \land z_l} \alpha_{\zbold} \ket{\psi_{\zbold}}  = (-1)^{\underset{(k, l) \in E_H}{\oplus} z'_k \land z'_l} \alpha_{\zbold'} \ket{\psi_{\zbold'}}.
\end{align} 
Recall that for complete graphs, all the honest vertices are connected to each other. Using this, we can simplify the exponent $\underset{(k, l) \in E_H}{\oplus} z_k \land z_l$. Each $(z_k \land z_l)$ will give 1 only if both $z_k, z_l$ are 1. So, we can rephrase this using the Hamming weight $\Delta(\zbold)$, which tells us how many 1s are in the string $\zbold$. Then, finding how many pairs $(z_k \land z_l)$ give 1 is equivalent to calculating $\Mycomb[\Delta(\zbold)]{2}$. 

Let us first take $\zbold, \zbold'$ such that $\Delta(\zbold), \Delta(\zbold') = 0$ (mod 2). We see that $(-1)^{\Mycomb[\Delta(\zbold)]{2}} = (-1)^{\Mycomb[\Delta(\zbold')]{2}}$ for $\Delta(\zbold)$ mod 4 = $\Delta(\zbold')$ mod 4. Similarly, if we take $\zbold, \zbold'$ such that $\Delta(\zbold), \Delta(\zbold') = 1$ (mod 2), we again find that $(-1)^{\Mycomb[\Delta(\zbold)]{2}} = (-1)^{\Mycomb[\Delta(\zbold')]{2}}$ for $\Delta(\zbold)$ mod 4 = $\Delta(\zbold')$ mod 4.
Thus, we have
\begin{align}
\forall \zbold, \zbold' \text{ such that } \Delta(\zbold) = 0 (\text{mod } 4), \Delta(\zbold') = 2 (\text{mod } 4), \alpha_{\zbold} \ket{\psi_{\zbold}} & = - \alpha_{\zbold'} \ket{\psi_{\zbold'}}, \nonumber \\
\forall \zbold, \zbold' \text{ such that } \Delta(\zbold) = 1 (\text{mod } 4), \Delta(\zbold') = 3 (\text{mod } 4), \alpha_{\zbold} \ket{\psi_{\zbold}} & = - \alpha_{\zbold'} \ket{\psi_{\zbold'}}.
\end{align}
We can now write our state as
\begin{align}
\ket{\Psi} = \ & \sum_{\zbold} \alpha_{\zbold} \ket{\zbold}_H \otimes \ket{\psi_{\zbold}}_D \nonumber \\
= \ & \alpha_{\zbold, \Delta(\zbold) = 0 (\text{mod } 2)} \Big[ \sum_{\Delta(\zbold) = 0 (\text{mod } 4)} \ket{\zbold} - \sum_{\Delta(\zbold) = 2 (\text{mod } 4)} \ket{\zbold} \Big]_H \otimes \ket{\psi_{\zbold, \Delta(\zbold) = 0 (\text{mod } 2)}}_D \nonumber \\
& + \alpha_{\zbold, \Delta(\zbold) = 1 (\text{mod } 2)} \Big[ \sum_{\Delta(\zbold) = 1 (\text{mod } 4)} \ket{\zbold} - \sum_{\Delta(\zbold) = 3 (\text{mod } 4)} \ket{\zbold} \Big]_H \otimes \ket{\psi_{\zbold, \Delta(\zbold) = 1 (\text{mod } 2)}}_D.
\label{eq:complcomp}
\end{align}
Then, our expressions for $\beta_{\xbold} \ket{\phi_{\xbold}}$ can be written using Equation (\ref{eq:yak}) as
\begin{align}
\beta_{00...00} \ket{\phi_{00...00}} & = \frac{\sqrt{2^{|H|}}}{2} \Big[ \alpha_{\zbold, \Delta(\zbold) = 0 (\text{mod } 2)} \ket{\psi_{\zbold, \Delta(\zbold) = 0 (\text{mod } 2)}} +  \alpha_{\zbold, \Delta(\zbold) = 1 (\text{mod } 2)} \ket{\psi_{\zbold, \Delta(\zbold) = 1 (\text{mod } 2)}} \Big], \nonumber \\
\beta_{11...11} \ket{\phi_{11...11}} & = \frac{\sqrt{2^{|H|}}}{2} \Big[ \alpha_{\zbold, \Delta(\zbold) = 0 (\text{mod } 2)} \ket{\psi_{\zbold, \Delta(\zbold) = 0 (\text{mod } 2)}} -   \alpha_{\zbold, \Delta(\zbold) = 1 (\text{mod } 2)} \ket{\psi_{\zbold, \Delta(\zbold) = 1 (\text{mod } 2)}} \Big].
\end{align}
Now, recall that $\bra{\phi_{00...00}}\ket{\phi_{11...11}} = 0$, and $\bra{\psi_{\Delta(\zbold) = 0(\text{mod } 2)}}\ket{\psi_{\Delta(\zbold) = 1(\text{mod } 2)}} = 0$. Using these orthogonality conditions and taking the inner product of the two expressions above, we get
\begin{align}
\abs{\alpha_{\zbold, \Delta(\zbold) = 0 (\text{mod } 2)}}^2 - \abs{\alpha_{\zbold, \Delta(\zbold) = 1 (\text{mod } 2)}}^2 = 0.
\end{align}
By normalisation, we have
\begin{align}
\abs{\alpha_{\zbold, \Delta(\zbold) = 0 (\text{mod } 2)}}^2 + \abs{\alpha_{\zbold, \Delta(\zbold) = 1 (\text{mod } 2)}}^2 = \frac{2}{2^{|H|}}.
\end{align}
Solving these, we find that $\alpha_{\zbold, \Delta(\zbold) = 0 (\text{mod } 2)}, \alpha_{\zbold, \Delta(\zbold) = 1 (\text{mod } 2)} = \pm \frac{1}{\sqrt{2^{|H|}}}$. Substituting this in Equation (\ref{eq:complcomp}) gives the state as 
\begin{align}
\ket{\Psi}
= \ & \pm \frac{1}{\sqrt{2^{|H|}}} \Bigg[ \Big( \sum_{\Delta(\zbold) = 0 (\text{mod } 4)} \ket{\zbold} - \sum_{\Delta(\zbold) = 2 (\text{mod } 4)} \ket{\zbold} \Big)_H \otimes \ket{\psi_{\zbold, \Delta(\zbold) = 0 (\text{mod } 2)}}_D \nonumber \\
& \pm \Big( \sum_{\Delta(\zbold) = 1 (\text{mod } 4)} \ket{\zbold} - \sum_{\Delta(\zbold) = 3 (\text{mod } 4)} \ket{\zbold} \Big)_H \otimes \ket{\psi_{\zbold, \Delta(\zbold) = 1 (\text{mod } 2)}}_D \Bigg],
\end{align}
which gives
\begin{align}
\rho_H^{\ket{\Psi}} = \ & \frac{1}{2^{|H|}} \Bigg[  \Big( \sum_{\Delta(\zbold) = 0 (\text{mod } 4)} \ket{\zbold} - \sum_{\Delta(\zbold) = 2 (\text{mod } 4)} \ket{\zbold} \Big)  \Big( \sum_{\Delta(\zbold) = 0 (\text{mod } 4)} \bra{\zbold} - \sum_{\Delta(\zbold) = 2 (\text{mod } 4)} \bra{\zbold} \Big) \nonumber \\
& + \Big( \sum_{\Delta(\zbold) = 1 (\text{mod } 4)} \ket{\zbold} - \sum_{\Delta(\zbold) = 3 (\text{mod } 4)} \ket{\zbold} \Big) \Big( \sum_{\Delta(\zbold) = 1 (\text{mod } 4)} \bra{\zbold} - \sum_{\Delta(\zbold) = 3 (\text{mod } 4)} \bra{\zbold} \Big) \Bigg] \nonumber \\
= \ & \rho_H^{\ket{\mathscr{G}}}.
\end{align}
Thus, $\ket{\Psi} = U_D \ket{\mathscr{G}}$.

\end{proof}

Inserting $\mathsf{J}=n$ into Lemmas \ref{lem:fullstablem} and \ref{lem:finalbit} concludes the proof.

\end{proof}

\subsection{Pentagon graph state}

The 5-qubit cycle graph state, in the shape of a pentagon, is known to be useful for secret sharing \cite{Markham}, as well as being the smallest quantum error correcting code that tolerates an arbitrary error on a single qubit \cite{Laflamme1996}. Let us now consider all possible sets of dishonest parties sharing such a graph (Figure \ref{fig:pentagon}), and see how we can reduce the resources required. We will show that in certain cases, by purely measuring the stabiliser generators, we can determine that if all tests pass perfectly, the state $\ket{\Psi} = U_D \ket{\mathscr{G}}$; this means that the parties can run the simpler version of Protocol \ref{alg:takeuchi} (with $\mathcal{S}$ as the set of generators). We summarise the result in Theorem \ref{th:graphpentproof}. 

\begin{manualtheorem}{2B}
If $\ket{\mathscr{G}}$ is a pentagon graph state with either one, three or four dishonest parties anywhere in the network, or two dishonest parties who are adjacent, and we set $N_{total} = 2n N_{test}$, $N_{test} = \lceil{ m n^4 \ln n}\rceil$, and $\mathcal{S}$ as the set of stabiliser generators, 
the probability that the fidelity of the averaged state of 
the target copy (over all possible choices of the tested copies and target copy) 
in Protocol \ref{alg:takeuchi} satisfies
\begin{align}
F(\rho_H^{\ket{\mathscr{G}}}, \rho_H^{avg})
\geq 1 - \frac{2\sqrt{c}}{n} - 2n \Big( 1 - \frac{N_{pass}}{n N_{test}} \Big) 
\end{align}
is at least $1 - n^{1 - \frac{2cm}{3}}$,
where 
$m, c$ are positive constants chosen such that $\frac{3}{2m} < c < \frac{(n-1)^2}{4}$.
\label{th:graphpentproof}
\end{manualtheorem}
\begin{proof}

We will tackle this proof taking all possible sets of dishonest parties separately. For the sets of dishonest parties specified in the statement of Theorem \ref{th:graphpentproof}, we will show certain steps in proving that by passing all stabiliser generator tests, our state must be $\ket{\Psi} = U_D \ket{\mathscr{G}}$. Then, as long as the honest parties know that the set of dishonest parties belongs to this `allowed set', they may simply measure the stabiliser generators instead of the full stabiliser group.

\subsubsection{One dishonest party}
We will start with a scenario where one party in the network is dishonest.
Without loss of generality, let us assume party 5 is dishonest, as shown in Figure \ref{fig:pentagon}(a). Then, we can write the Schmidt decomposition of the ideal state $\ket{\mathscr{G}}$ with respect to the partition $(H, D)$ as
\begin{align}
\ket{\mathscr{G}} = \ & \frac{1}{\sqrt{2^{4}}} \sum_{\zbold} (-1)^{(z_1 \land z_2) \oplus (z_2 \land z_3) \oplus (z_3 \land z_4)} \ket{\zbold}_H \otimes (\sigma_Z^{(5)})^{z_1} (\sigma_Z^{(5)})^{z_4} \ket{\mathscr{G-H}}_D \nonumber \\
= \ & \frac{1}{4} \Big[ (\ket{0000} + \ket{1001} + \ket{0010} - \ket{1011} + \ket{0100} - \ket{1101} - \ket{0110} - \ket{1111})_H \otimes \ket{\mathscr{G-H}_{(0)}}_D \nonumber \\
& + ( \ket{0001} + \ket{1000} - \ket{0011} + \ket{1010} + \ket{0101} - \ket{1100} + \ket{0111} + \ket{1110})_H \otimes \ket{\mathscr{G-H}_{(1)}}_D \Big].
\end{align}
(Note that, as expected, the dishonest part of the state, $\ket{\mathscr{G-H}}$, is the same for honest parts with the same value of $\underset{i \in N(d)}{\oplus} z_i = z_1 \oplus z_4$.) 

\begin{figure*}[t]
\centering
\includegraphics[trim = 2cm 0cm 7cm 1cm, width=0.6\textwidth]{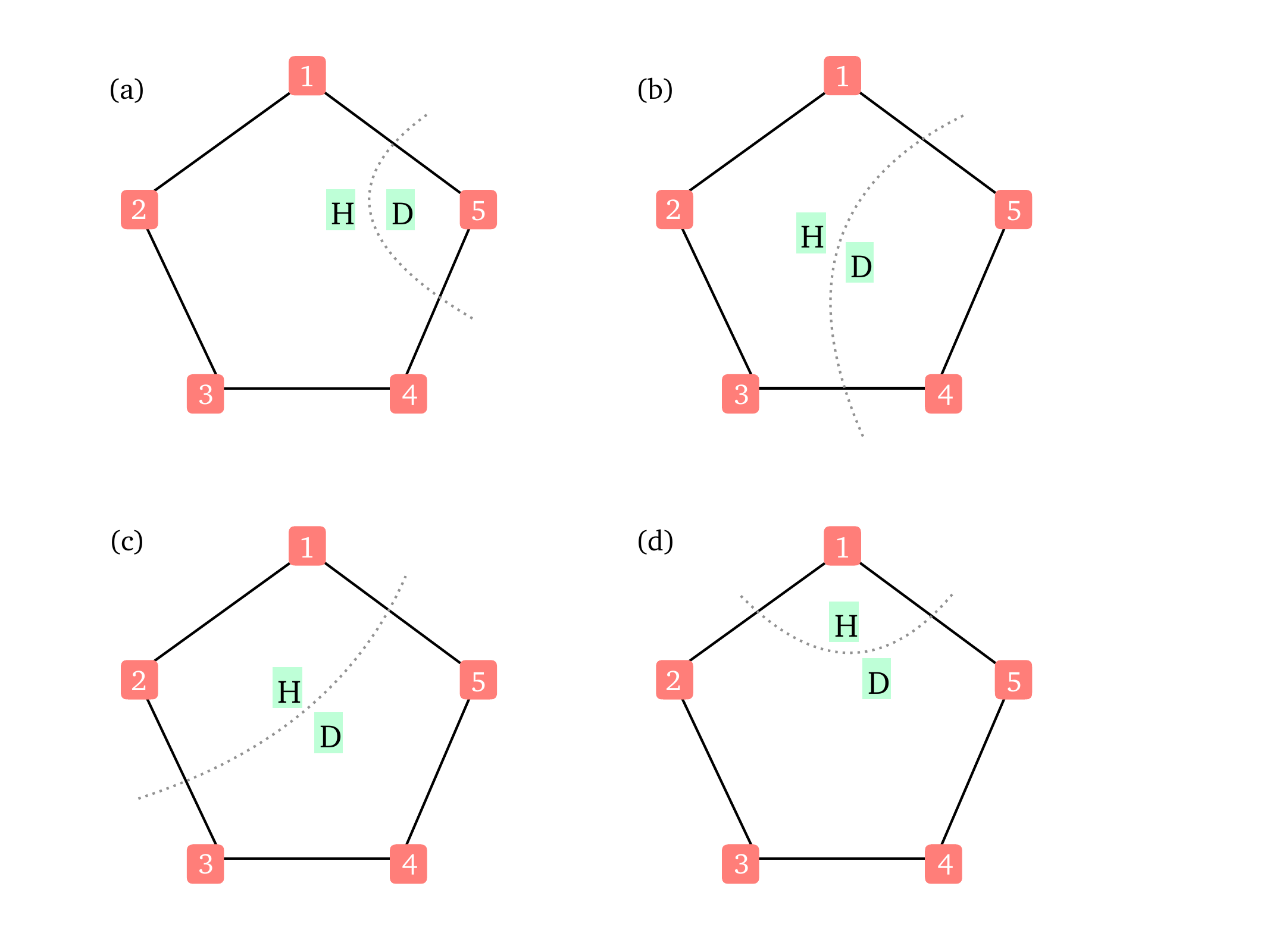}
\caption{Sets of honest (H) and dishonest (D) parties for the pentagon graph state considered in Theorem \ref{th:graphpentproof}.}
\label{fig:pentagon}
\end{figure*}

As usual, we start by writing our state in the honest computational and subgraph bases as in Equations (\ref{eq:compbas}) and (\ref{eq:subbas}).
Now, the test measurements, or the stabiliser generators in this case, are given by $\{ XZ\mathds{1}\mathds{1}Z'$, $ ZXZ\mathds{1}\mathds{1}'$, $ \mathds{1}ZXZ\mathds{1}'$, $\mathds{1}\mathds{1}ZXZ'$, $ Z\mathds{1}\mathds{1}ZX' \}$. We group them into Group 1 measurements $\{ Z\mathds{1}\mathds{1}ZX'\}$ and Group 2 measurements $\{ XZ\mathds{1}\mathds{1}Z', ZXZ\mathds{1}\mathds{1}', \mathds{1}ZXZ\mathds{1}', \mathds{1}\mathds{1}ZXZ' \}$. 
From the Group 1 measurement passing perfectly, we see that $\forall \zbold, \zbold'$ such that $z_1 \oplus z_4 \neq z_1' \oplus z_4'$, we must have $\bra{\psi_{\zbold}}\ket{\psi_{\zbold'}} = 0$. 
(Note that this matches Equation (\ref{eq:psiorth}).) 
From the Group 2 measurements passing perfectly, we know that $\forall \xbold, \xbold'$, we have $\bra{\phi_{\xbold}}\ket{\phi_{\xbold'}} = 0$.

We will now see how to set certain $\beta_{\xbold} = 0$ from just the stabiliser generators alone. The measurements $ZXZ\mathds{1}\mathds{1}', \mathds{1}ZXZ\mathds{1}'$ must give outcome $+1$ to pass the test. Since the dishonest party asked to measure $\mathds{1}'$ will always output $+1$, the honest measurements $ZXZ\mathds{1}, \mathds{1}ZXZ$ must also give outcome $+1$. Further, the measurements $XZ\mathds{1}\mathds{1}Z', \mathds{1}\mathds{1}ZXZ'$ give outcome $+1$. Since the dishonest party asked to measure $Z'$ does not know whether the honest parties are measuring $XZ\mathds{1}\mathds{1}$ or $\mathds{1}\mathds{1}ZX$, yet still manages to make the test pass perfectly, this means that the honest part of the state must give the same outcome for these two measurements. With these conditions, we have $\beta_{\xbold} = 0$ $ \forall \xbold \notin \{0000, 1001 \}$. (Note that we get the same result from the full stabiliser group: 
the possible sets $A$ are $ \{ 1, 4 \}, \{ 2 \}, \{ 3 \}, \{ 2, 3 \}, \{ 1, 2, 4 \}, \{ 1, 3, 4 \}, \{ 1, 2, 3, 4 \}$ and so $\beta_{\xbold} = 0$ if $x_1 \oplus x_4 = 1, x_2 = 1$, or $x_3 = 1$.)

Following the remaining steps of the general proof, we then get
\begin{align}
\ket{\Psi} = \ & \pm \frac{1}{\sqrt{2^{4}}} \Big[ (\ket{0000} + \ket{1001} + \ket{0010} - \ket{1011} + \ket{0100} - \ket{1101} - \ket{0110} - \ket{1111})_H \otimes \ket{\phi_{0000}}_D \nonumber \\
& \pm ( \ket{0001} + \ket{1000} - \ket{0011} + \ket{1010} + \ket{0101} - \ket{1100} + \ket{0111} + \ket{1110})_H \otimes \ket{\phi_{1001}}_D \Big],
\end{align}
which gives $\ket{\Psi} = U_D \ket{\mathscr{G}}$. 

\subsubsection{Two dishonest parties}

Let us start by considering parties 1, 2 and 3 as honest and parties 4 and 5 as dishonest, pictured in Figure \ref{fig:pentagon}(b). We will vary the dishonest parties later and see how this affects the results, but for now let us consider the two dishonest parties to be adjacent to one another. The Schmidt decomposition of the ideal graph state for this partition is given by
\begin{align}
\ket{\mathscr{G}} = \ & \frac{1}{\sqrt{2^{3}}} \sum_{\zbold} (-1)^{(z_1 \land z_2) \oplus (z_2 \land z_3)} \ket{\zbold}_H \otimes (\sigma_Z^{(5)})^{z_1} (\sigma_Z^{(4)})^{z_3} \ket{\mathscr{G-H}}_D \nonumber \\
= \ & \frac{1}{2\sqrt{2}} \Big[ (\ket{000} + \ket{010})_H \otimes \ket{\mathscr{G-H}_{(00)}}_D + (\ket{001} - \ket{011})_H \otimes \ket{\mathscr{G-H}_{(10)}}_D \nonumber \\
& + (\ket{100} - \ket{110})_H \otimes \ket{\mathscr{G-H}_{(01)}}_D + (\ket{101} + \ket{111})_H \otimes \ket{\mathscr{G-H}_{(11)}}_D \Big].
\end{align}
(It is easy to see that for states with the same $\underset{i \in N(d)}{\oplus} z_i$, $\forall d \in D$, which here refers to states with the same value of $z_1$ and $z_3$, the corresponding dishonest part $\ket{\mathscr{G-H}}$ is the same.)

Let us now see whether we can make the desired statement by purely considering the stabiliser generators,
$
\{ XZ\mathds{1}(\mathds{1}Z)', ZXZ(\mathds{1}\mathds{1})', \mathds{1}ZX(Z\mathds{1})', \mathds{1}\mathds{1}Z(XZ)', Z\mathds{1}\mathds{1}(ZX)' \}
$.
We group them, as usual, into Group 1 given by $\{ \mathds{1}\mathds{1}Z(XZ)', Z\mathds{1}\mathds{1}(ZX)' \}$, and Group 2 given by $\{ XZ\mathds{1} (\mathds{1}Z)', ZXZ(\mathds{1}\mathds{1})', \mathds{1}ZX(Z\mathds{1})' \}$. By the measurements in Group 1 passing perfectly, we see that $\forall \zbold, \zbold'$ such that $z_1 \neq z_1'$ or $z_3 \neq z_3'$, we have $\bra{\psi_{\zbold}}\ket{\psi_{\zbold'}} = 0$, which can also be seen from Equation (\ref{eq:psiorth}).
By the measurements in Group 2 passing perfectly, we have $\forall \xbold, \xbold'$, $\bra{\phi_{\xbold}}\ket{\phi_{\xbold'}} = 0$.

Now, Group 2 contains the measurement $ZXZ(\mathds{1}\mathds{1})'$. In order to get outcome $+1$ here, the only terms of the honest subgraph that can appear in Equation (\ref{eq:subbas}) are those that give $+1$ when measuring $ZXZ$. Thus, the Group 2 measurements passing perfectly tells us that $\beta_{\xbold} = 0$ $\forall \xbold \notin \{ 000, 001, 100, 101 \}$. (Note that by using the full set of stabiliser measurements, we do not get more information than this, as the only possible set $A = \{ 2 \}$ and so we know that if $x_2 = 1$, then $\beta_{\xbold} = 0$.) Continuing with the steps of the general proof, we obtain $\ket{\Psi} = U_D \ket{\mathscr{G}}$. 

By inspecting the stabiliser generators, we see that such an analysis holds whenever the two dishonest parties are adjacent, as there will always be a measurement $(\mathds{1}\mathds{1})'$ that forces the corresponding honest outcome to be $+1$. Thus, in this case, it suffices to measure the stabiliser generators. 
If the two dishonest parties are not adjacent, however, there will not be such a measurement $(\mathds{1}\mathds{1})'$, and so the parties must measure the full set of stabilisers. 

\subsubsection{Three dishonest parties}

Let us now assume parties 1 and 2 are honest, while parties 3, 4 and 5 are dishonest, as in Figure \ref{fig:pentagon}(c), to start with. The Schmidt decomposition is given by
\begin{align}
\ket{\mathscr{G}} = \ & \frac{1}{\sqrt{2^2}} \sum_{\zbold} (-1)^{z_1 \land z_2} \ket{\zbold}_H \otimes (\sigma_Z^{(5)})^{z_1} (\sigma_Z^{(3)})^{z_2} \ket{\mathscr{G-H}}_D \nonumber \\
= \ & \frac{1}{2} \Big[ \ket{00}_H \otimes \ket{\mathscr{G-H}_{(000)}}_D + \ket{01}_H \otimes \ket{\mathscr{G-H}_{(100)}}_D \nonumber \\
& \ \ \ + \ket{10}_H \otimes \ket{\mathscr{G-H}_{(001)}}_D - \ket{11}_H \otimes \ket{\mathscr{G-H}_{(101)}}_D \Big].  
\end{align}
(Here, as in the general proof, we see that both $z_1$ and $z_2$ must be the same in order for the corresponding $\ket{\mathscr{G-H}}$ to be the same.)

As we see here, there is no grouping of $\ket{\zbold}$ terms, and so there will be no $\beta_{\xbold} = 0$. 
The Group 1 measurements are then $\{ \mathds{1}Z(XZ\mathds{1})', \mathds{1}\mathds{1}(ZXZ)', Z\mathds{1}(\mathds{1}ZX)' \}$, and the Group 2 measurements are $\{ XZ(\mathds{1}\mathds{1}Z)', ZX(Z\mathds{1}\mathds{1})' \}$. 
We see that if $z_1 \neq z_1'$ or $z_2 \neq z_2'$, the respective states are orthogonal. This implies that $\bra{\psi_{\zbold}}\ket{\psi_{\zbold'}} = 0$ $\forall \zbold, \zbold'$.
We also have $\forall \xbold, \xbold'$, $\bra{\phi_{\xbold}}\ket{\phi_{\xbold'}} = 0$. From passing the Group 2 measurements, we cannot set any $\beta_{\xbold} = 0$ (using the full stabiliser set, we see that $A$ is the empty set). Then, we simply follow the steps of the general proof to get $\ket{\Psi} = U_D \ket{\mathscr{G}}$. 
This reasoning holds for any two parties being dishonest, no matter whether they are adjacent, and so for this case, the parties only need to measure the stabiliser generators. 

\subsubsection{Four dishonest parties}

In this scenario where only one party is honest, the analysis is again simple. 
Without loss of generality, let us assume only party 1 is honest, as shown in Figure \ref{fig:pentagon}(d). First, let us write the Schmidt decomposition for this partition $(H, D)$ as
\begin{align}
\ket{\mathscr{G}} & = \frac{1}{\sqrt{2^{1}}} \sum_{\zbold} \ket{\zbold}_H \otimes (\sigma_Z^{(2)} \sigma_Z^{(5)})^{z_1} \ket{\mathscr{G-H}}_D \nonumber \\
& = \frac{1}{\sqrt{2}} \Big[ \ket{0}_H \otimes \ket{\mathscr{G-H}_{(0000)}}_D + \ket{1}_H \otimes \ket{\mathscr{G-H}_{(1001)}}_D \Big].
\end{align}
(Note that, as in the general proof, if $\underset{i \in N(d)}{\oplus} z_i = z_1$ is the same, the corresponding $\ket{\mathscr{G-H}}$ is the same.)

Now, the set of measurements is given by the stabiliser generators again, which we will separate into Group 1 containing $\{ Z(XZ \mathds{1} \mathds{1})',  \mathds{1}(ZXZ \mathds{1})',  \mathds{1}( \mathds{1}ZXZ)', Z( \mathds{1} \mathds{1}ZX)' \}$, and Group 2 containing $\{ X(Z \mathds{1} \mathds{1}Z)' \}$. Group 1 passing perfectly gives $\bra{\psi_0}\ket{\psi_1} = 0$, since $\ket{0}_H, \ket{1}_H$ will give different outcomes for $Z$, and so the dishonest parties must be able to perfectly discriminate between their corresponding parts of the state in order to pass perfectly (we reach the same conclusion using Equation (\ref{eq:psiorth})). 
Similarly, Group 2 passing perfectly gives $\bra{\phi_0}\ket{\phi_1} = 0$, since $\ket{+}_H, \ket{-}_H$ give different outcomes for $X$. 
When there is only one honest party, we cannot set any $\beta_{\xbold} = 0$ from passing perfectly (the set $A$ is the empty set, so this gives the same result as measuring the full stabiliser set). 
We then proceed with the proof to get $\ket{\Psi} = U_D \ket{\mathscr{G}}$.

Thus, for any number of adjacent dishonest parties, passing the stabiliser generator tests of Protocol \ref{alg:takeuchi} allows us to conclude that the state in each round is $\ket{\Psi} = U_D \ket{\mathscr{G}}$. Inserting $\mathsf{J}=n$ into  Lemmas \ref{lem:fullstablem} and \ref{lem:finalbit} concludes the proof.

\end{proof}

\subsection{Cycle graph states}

With the example of the pentagon graph state in the previous subsection, we notice some features which can be extended to general $n$-qubit cycle (or ring) graph states (Figure \ref{fig:cycle}), used in various applications such as quantum error correction \cite{Gottesman1997} and quantum walks \cite{Aharonov2001}.
In Theorem \ref{th:graphcycleproof}, we give some cases for the cycle graph state where it suffices to measure only stabiliser generators in Protocol \ref{alg:takeuchi}. We will outline the reasoning behind this, although it can be seen explicitly in the pentagon graph state example.

\begin{manualtheorem}{2C}
If $\ket{\mathscr{G}}$ is a cycle graph state with either one, $n-2$ or $n-1$ dishonest parties anywhere in the network, or any other number of dishonest parties who are adjacent, and we set $N_{total} = 2n N_{test}$, $N_{test} = \lceil{ m n^4 \ln n}\rceil$, and $\mathcal{S}$ as the set of stabiliser generators, 
the probability that the fidelity of the averaged state of 
the target copy (over all possible choices of the tested copies and target copy) 
in Protocol \ref{alg:takeuchi} satisfies
\begin{align}
F(\rho_H^{\ket{\mathscr{G}}}, \rho_H^{avg}) 
\geq 1 - \frac{2\sqrt{c}}{n} - 2n \Big( 1 - \frac{N_{pass}}{n N_{test}} \Big) 
\end{align}
is at least $1 - n^{1 - \frac{2cm}{3}}$, 
where 
$m, c$ are positive constants chosen such that $\frac{3}{2m} < c < \frac{(n-1)^2}{4}$.
\label{th:graphcycleproof}
\end{manualtheorem}

\begin{proof-sketch}
Whenever there are $n-1$ or $n-2$ dishonest parties in the network, the Schmidt decomposition is given by
\begin{align}
\ket{\mathscr{G}} = \frac{1}{\sqrt{2^{|H|}}} \sum_{\zbold} \ket{\zbold}_H \otimes \ket{\mathscr{G-H}_{(\zbold)}}_D,
\end{align}
where each $\ket{\zbold}$ corresponds to a different dishonest part $\ket{\mathscr{G-H}}$. This means that there is no grouping of $\ket{\zbold}$ terms, and so the Group 2 measurements do not tell us that any $\beta_{\xbold} = 0$, as we saw in the previous example. Thus, in this case the parties can simply measure the stabiliser generators (the full stabiliser group would give no additional information, as the set $A$ is empty).

\begin{figure*}[t]
\centering
\includegraphics[trim = 0cm 5.5cm 0cm 7cm, width=0.7\textwidth]{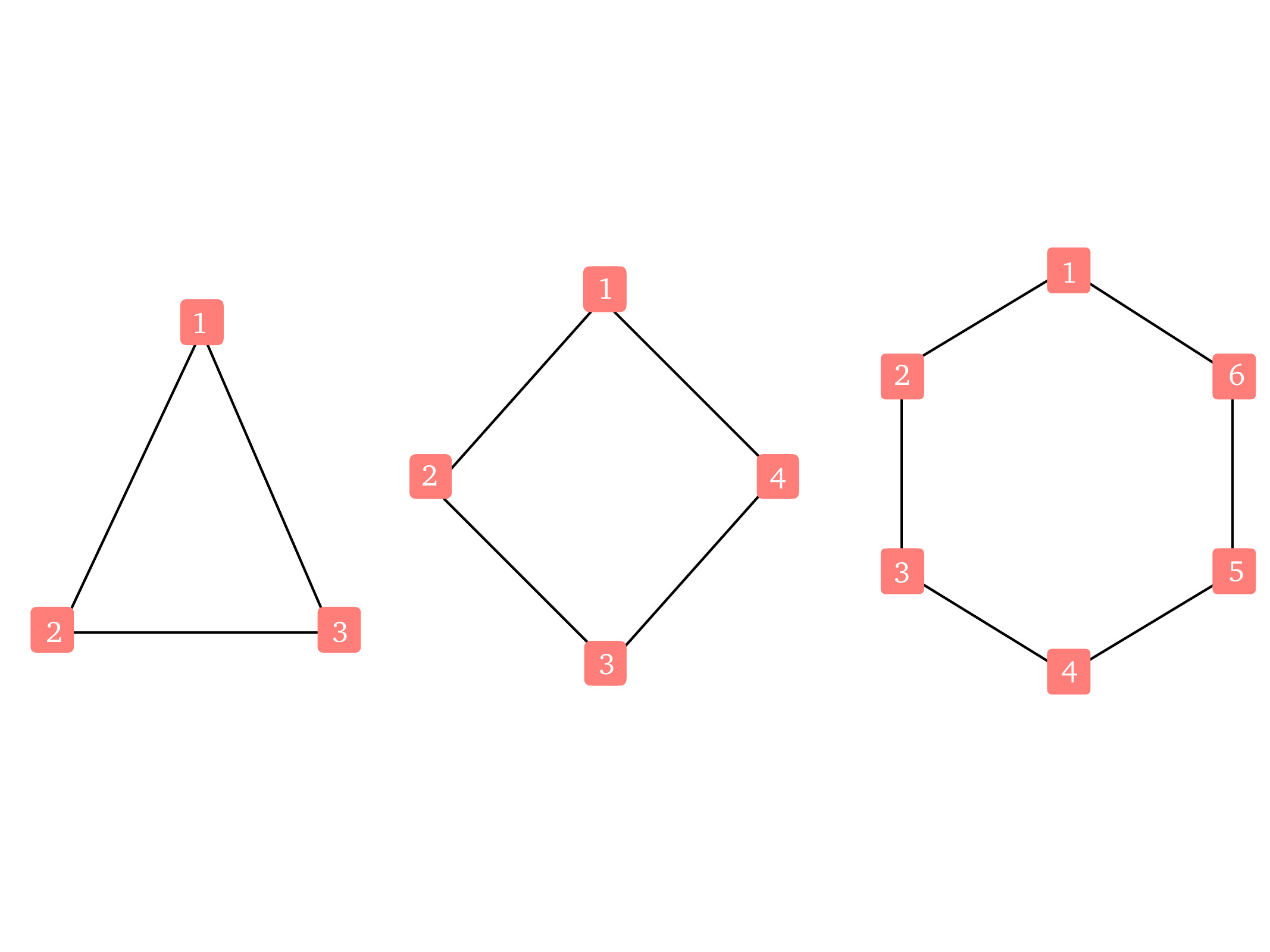}
\caption{3-qubit, 4-qubit and 6-qubit cycle (ring) graph states.}
\label{fig:cycle}
\end{figure*}

Let us now consider the case of one dishonest party.
The dishonest measurements of $\mathds{1}'$, corresponding to the generators of qubits $i \notin N(d)$, ensure that if $x_{i \notin N(d)} = 1$, then $\beta_{\xbold} = 0$. Further, the two dishonest measurements of $Z'$, corresponding to the generators of the two qubits $i \in N(d)$, ensure that if $x_{i \in N(d)}$ are not the same (in order to give the same honest outcome), then $\beta_{\xbold} = 0$.
This means that the only terms in the honest subgraph are $\ket{\mathscr{H}_{(00...00)}}$ and $\ket{\mathscr{H}_{(10...01)}}$, which allows us to arrive at our result from just the stabiliser generators. (Note that using the full stabiliser group, Equation (\ref{eq:condition}) would imply that if $\underset{i \in N(d)}{\oplus} x_i = 1$, or $x_{i \notin N(d)} = 1$, then $\beta_{\xbold} = 0$, which gives the same result.)

Finally, for any other number of honest/dishonest parties, if the dishonest parties are adjacent then there will be measurements $(\mathds{1}...\mathds{1})'$ in the set of stabiliser generators, corresponding to the generators of honest vertices that are not connected to any of the dishonest vertices (for qubits $i \notin N(D)$). Since the dishonest part must give outcome $+1$ for this measurement, the honest part must also give outcome $+1$ in order to pass perfectly, any terms where $x_{i \notin N(D)} = 1$ must have $\beta_{\xbold} = 0$. (Note that using the full stabiliser group, the possible sets $A$ are all combinations of honest vertices that are not connected to any dishonest vertex, which gives the same result.)
Thus, we can conclude that $\ket{\Psi} = U_D \ket{\mathscr{G}}$ from only measuring the stabiliser generators. 

Inserting $\mathsf{J}=n$ into Lemmas \ref{lem:fullstablem} and \ref{lem:finalbit} concludes the proof.

\end{proof-sketch}

In fact, there are additional examples that do not fall into these categories, but where it suffices to measure the stabiliser generators only. For example, consider the 6-qubit cycle graph state with $D = \{ 2, 3, 6 \}$, the 7-qubit cycle graph state with $D = \{ 2, 3, 4, 7 \}$, or the 8-qubit cycle graph state with $D = \{ 2, 3, 6, 7 \}$. There are no $\beta_{\xbold} = 0$ in the Schmidt decomposition for these cases, which means that by passing the stabiliser generator measurements in Protocol \ref{alg:takeuchi} perfectly, we can determine that $\ket{\Psi} = U_D \ket{\mathscr{G}}$.

\subsection{Cluster states}
Cluster states correspond to lattices of dimension $\mathsf{D}$, and have been shown to be a central resource in MBQC \cite{Raussendorf2001, Raussendorf2003}. 
For verification of such states in an untrusted network, we will now give some examples of scenarios where we can reduce the number of test measurements to purely the stabiliser generators (thus allowing the parties to run the simpler verification scheme in Protocol \ref{alg:takeuchi}) with a smaller set of test measurements. Again, these are not the only possible sets of dishonest parties in a cluster state network that allow such a simplification; there may be many more examples.

Linear, or one-dimensional, cluster states (also known as path graph states) correspond to qubits connected in a line (Figure \ref{fig:path}). In Theorem \ref{th:graphpathproof}, we give some cases where it is sufficient to measure the stabiliser generators. 

\begin{manualtheorem}{2D}
If $\ket{\mathscr{G}}$ is a $\mathsf{1D}$ cluster state with either one or $n-1$ dishonest parties anywhere in the network, or any other number of adjacent honest or dishonest parties, and we set $N_{total} = 2n N_{test}$, $N_{test} = \lceil{ m n^4 \ln n}\rceil$, and $\mathcal{S}$ is the set of stabiliser generators, 
the probability that the fidelity of the averaged state of 
the target copy (over all possible choices of the tested copies and target copy) 
in Protocol \ref{alg:takeuchi} satisfies
\begin{align}
F(\rho_H^{\ket{\mathscr{G}}}, \rho_H^{avg}) 
\geq 1 - \frac{2\sqrt{c}}{n} - 2n \Big( 1 - \frac{N_{pass}}{n N_{test}} \Big) 
\end{align}
is at least $1 - n^{1 - \frac{2cm}{3}}$,
where 
$m, c$ are positive constants chosen such that $\frac{3}{2m} < c < \frac{(n-1)^2}{4}$.
\label{th:graphpathproof}
\end{manualtheorem}
\begin{proof}

The stabiliser generators of an $n$-qubit $\mathsf{1D}$ cluster state are given by
$
\{ XZ\mathds{1}...\mathds{1}$, $ZXZ\mathds{1}...\mathds{1}$, $..., \mathds{1}...\mathds{1}ZX \}
$.
As we noted for cycle graphs, if there are $n-1$ dishonest parties in the network, it suffices to measure only stabiliser generators, as passing the Group 2 measurements perfectly does not tell us to set any $\beta_{\xbold} = 0$. 

If there is one dishonest party at either end of the line, we see that to get overall outcome $+1$, the honest part of the stabiliser generator measurements corresponding to every honest vertex $i \notin N(d)$ must give outcome $+1$. This means that if $x_{i \notin N(d)} = 1$, we must have $\beta_{\xbold} = 0$. (Note that from the full stabiliser group, we see that all possible sets $A$ contain only honest vertices $i \notin N(d)$, and so we come to the same conclusion.) If the dishonest party is not at the end of the line, but at a vertex connected to two other (honest) vertices, we again must have that if $x_{i \notin N(d)} = 1$ then $\beta_{\xbold} = 0$, and further, the honest part of the stabiliser generators for $i \in N(d)$ must give the same outcome (as they have the same dishonest part). This means that if $x_i$ does not have the same value for both $i \in N(d)$, then $\beta_{\xbold} = 0$. 
(Note that this gives the same result as using the full stabiliser group, as the set $A$ may consist of any combination of honest vertices $i \notin N(d)$ as well as both honest vertices $i \in N(d)$.)


\begin{figure*}[t]
\centering
\includegraphics[trim = 0cm 12cm 0cm 12cm, width=0.7\textwidth]{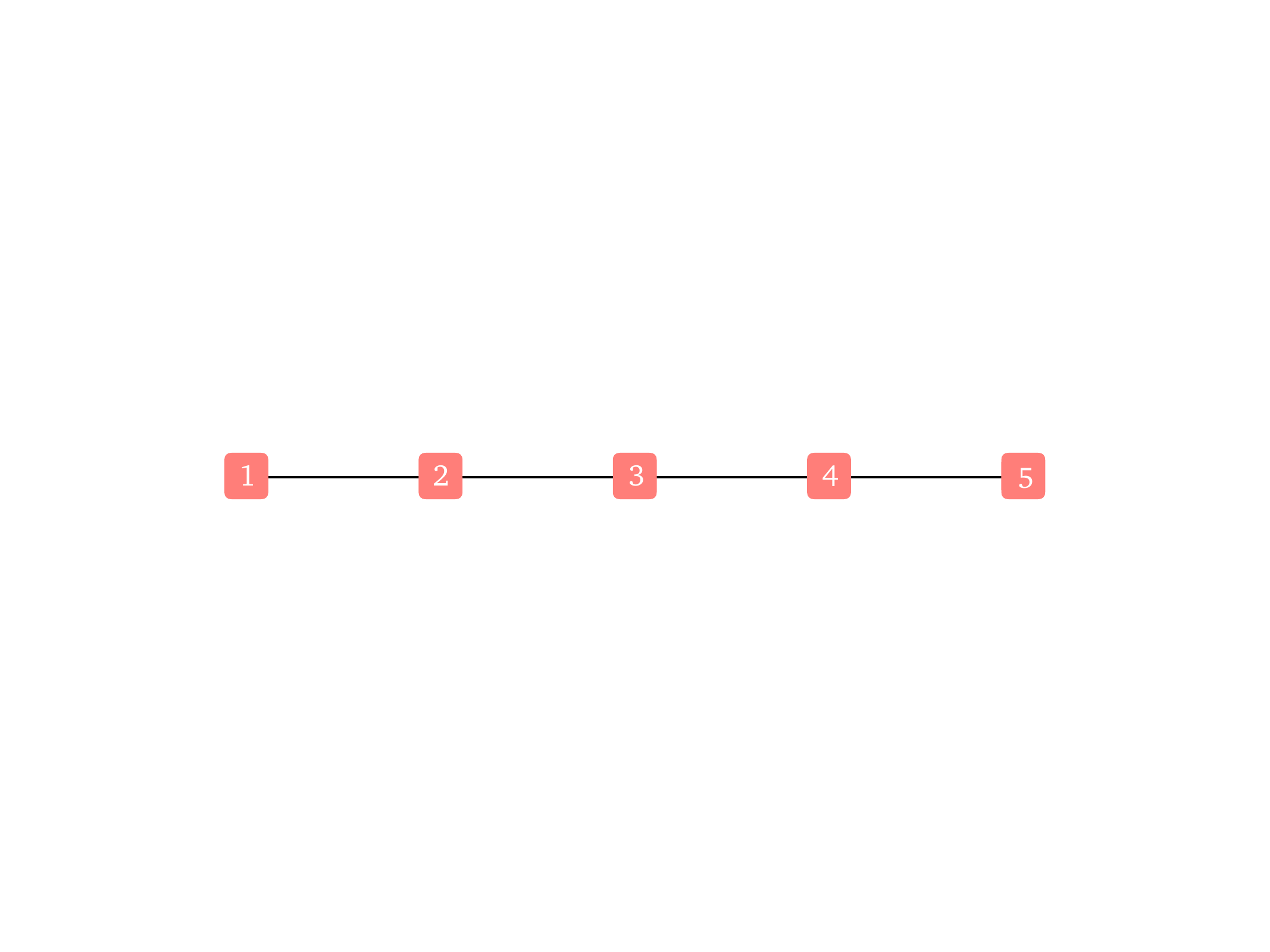}
\caption{5-qubit path graph state, or $\mathsf{1D}$ cluster state.}
\label{fig:path}
\end{figure*}

Further, for any other number of dishonest parties, if it is known that the honest parties are all adjacent to each other, or the dishonest parties are all adjacent to each other, then the parties could measure only the stabiliser generators. In such a setting, there will be $(\mathds{1}...\mathds{1})'$ measurements in the stabiliser generators (corresponding to $i \notin N(D)$) that allow us to set $\beta_{\xbold} = 0$ if $x_{i \notin N(D)} = 1$. (Note that from the full stabiliser group, all possible sets $A$ will consist of $i \notin N(D)$, and so this gives the same result.) This corresponds to the first $|H|$ qubits of the line belonging to the honest parties, and the remaining $|D|$ qubits belonging to the dishonest parties (or vice versa).

Inserting $\mathsf{J}=n$ into Lemmas \ref{lem:fullstablem} and \ref{lem:finalbit} concludes the proof.

\end{proof}

Additional examples of path graph states where it is sufficient to measure only the stabiliser generators include, for the 5-qubit path graph, the sets $D = \{ 1, 2, 5 \}$, or $D = \{ 1, 3, 5 \}$, where there are no $\beta_{\xbold} = 0$ in the Schmidt decomposition, and so measuring the stabiliser generators suffices.

Two-dimensional cluster states have the underlying structure of a $t \times t$ square grid, where the total number of qubits in the lattice is $n = t \times t$. In Theorem \ref{th:graphclusterproof}, we give certain sets of dishonest parties for which it is possible to verify such states with the resource-efficient Protocol \ref{alg:takeuchi} requiring only generator tests. This is a useful result for verification in the MBQC paradigm.

\begin{manualtheorem}{2E}
If $\ket{\mathscr{G}}$ is a $\mathsf{2D}$ cluster state with either one or $n-1$ dishonest parties anywhere in the network, or any other set of adjacent dishonest parties that
forms a square or rectangle anywhere in the network,
and we set $N_{total} = 2n N_{test}$, $N_{test} = \lceil{ m n^4 \ln n}\rceil$, and $\mathcal{S}$ as the set of stabiliser generators, 
the probability that the fidelity of the averaged state of 
the target copy (over all possible choices of the tested copies and target copy) 
in Protocol \ref{alg:takeuchi} satisfies
\begin{align}
F(\rho_H^{\ket{\mathscr{G}}}, \rho_H^{avg}) 
\geq 1 - \frac{2\sqrt{c}}{n} - 2n \Big( 1 - \frac{N_{pass}}{n N_{test}} \Big) 
\end{align}
is at least $1 - n^{1 - \frac{2cm}{3}}$, 
where 
$m, c$ are positive constants chosen such that $\frac{3}{2m} < c < \frac{(n-1)^2}{4}$.
\label{th:graphclusterproof}
\end{manualtheorem}

\begin{figure*}[t]
\centering
\includegraphics[trim = 4cm -1cm 4cm 0cm, width=0.6\textwidth]{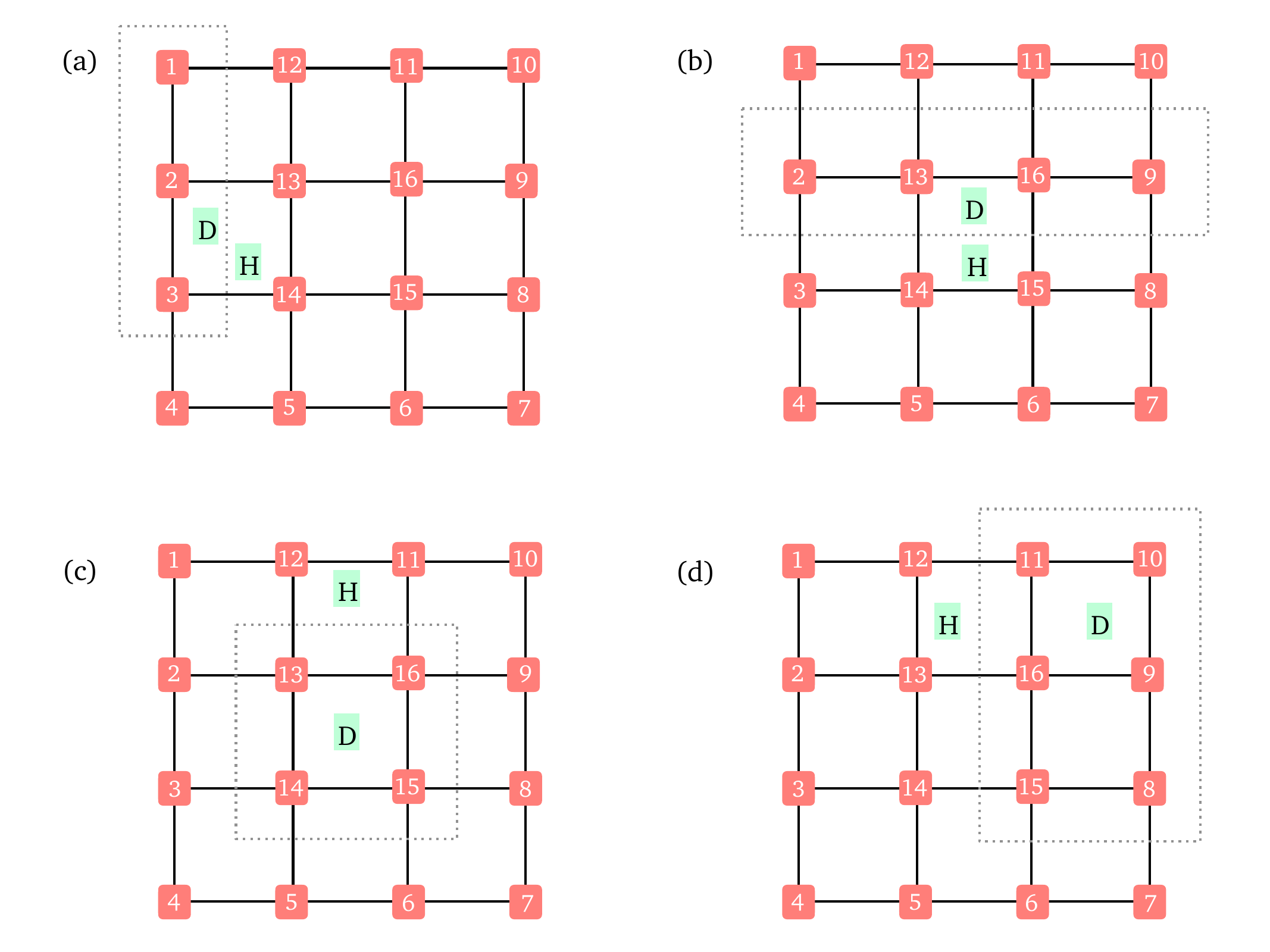}
\caption{Sets of honest (H) and dishonest (D) parties for the $4 \times 4$ cluster state considered in Theorem \ref{th:graphclusterproof}.}
\label{fig:cluster}
\end{figure*}

\begin{proof}
Similarly to what we have seen in previous examples, if there is one honest party, or one dishonest party, it is enough to measure only the stabiliser generators. We will now show that, for any number of dishonest parties in a $\mathsf{2D}$ cluster state, if they form a square or rectangle in the lattice, it is possible to do the verification using only the generators. 

Consider the example of a $4 \times 4$ (or 16-qubit) cluster state, with example sets of honest and dishonest parties as shown in Figure \ref{fig:cluster}. Let us start with the parties 1, 2 and 3 being dishonest, while the remaining are honest, as in Figure \ref{fig:cluster}(a). If we write down the set of stabiliser generators, we find that the generators corresponding to qubits 5, 6, 7, 8, 9, 10, 11, 15 and 16 will have dishonest part $(\mathds{1}\mathds{1}\mathds{1})'$, as they are not connected to the dishonest set. This means that the corresponding honest measurements on the state $\ket{\Psi}$ must give outcome $+1$. Recall that the generator of qubit $i$ contains the measurement $X_i$, which tells us that if $x_i = 1$ for $i \in \{ 5, 6, 7, 8, 9, 10, 11, 15, 16 \}$ (or equivalently, if $x_{i \notin N(D)} = 1$), then $\beta_{\xbold} = 0$. Further, the dishonest part of the measurement will be the same for generators of qubits 4 and 14 (since they are both connected to dishonest qubit 3). This means that in order to pass these test measurements, the corresponding honest part of the measurements must give the same outcome, and so we must have that if $x_4 \neq x_{14}$, then $\beta_{\xbold} = 0$.  (Note that from the full stabiliser group, we have all possible sets $A$ containing $i \notin N(D)$ as well as both $i = 4, 14$ which are evenly connected to all $d \in D$. This gives the same conclusions as previously, and so it is clear that only measuring the generators is necessary.)

Now, consider the example shown in Figure \ref{fig:cluster}(b), where parties 2, 13, 16 and 9 are dishonest, while the remaining are honest. Again, writing down the set of stabiliser generators, we find that the generators corresponding to qubits 4, 5, 6 and 7 have dishonest part $(\mathds{1}\mathds{1}\mathds{1}\mathds{1})'$, which tells us that if $x_i = 1$ for $i \in \{ 4, 5, 6, 7 \}$, then $\beta_{\xbold} = 0$. Further, the generators of qubits 1 and 3, 12 and 14, 11 and 15, and 8 and 10 have the same dishonest part, and so we must have that if either $x_1 \neq x_3, x_{12} \neq x_{14}, x_{11} \neq x_{15}$, or $x_8 \neq x_{10}$, then $\beta_{\xbold} = 0$. 
(Note that from the full stabiliser group, the set $A$ consists of all combinations of $i \notin N(D)$, as well as combinations of the pairs of vertices $(1, 3), (12, 14), (11, 15), (8, 10)$ which are evenly connected to all $d \in D$, leading to the same result.)

Let us now move on to the example shown in Figure \ref{fig:cluster}(c), where parties 13, 14, 15 and 16 are dishonest, and the remaining are honest. The generators of qubits 1, 4, 7 and 10 have dishonest part $(\mathds{1}\mathds{1}\mathds{1}\mathds{1})'$, which means that if $x_i = 1$ for $i \in \{ 1, 4, 7, 10 \}$, then $\beta_{\xbold} = 0$. Further, the generators of qubits 1 and 12, 9 and 11, 3 and 5, and 6 and 8 have the same dishonest part, which means that additionally, if either $x_1 \neq x_{12}, x_9 \neq x_{11}, x_3 \neq x_5$, or $x_6 \neq x_8$, then $\beta_{\xbold} = 0$. (Note that from the full stabiliser group, we come to the same conclusions, as the set $A$ contains all combinations of $i \notin N(D)$ as well as the pairs $(1, 12), (9, 11), (3, 5), (6, 8)$.) 

As another example, consider the dishonest set to comprise of parties 8, 9, 10, 11, 15 and 16, as shown in Figure \ref{fig:cluster}(d). From the generators whose dishonest part is $(\mathds{1}...\mathds{1})'$, we must have that if $x_i = 1$ for $i \in \{ 1, 2, 3, 4, 5 \}$, then $\beta_{\xbold} = 0$. Further, from the generators whose dishonest part is the same, we have that if $x_6 \neq x_{14}$, then $\beta_{\xbold} = 0$. (Note that from the full stabiliser group, we see that the set $A$ can contain all combinations of $i \notin N(D)$ along with the pair $(6, 14)$.)

Such an analysis easily extends to all cases where the dishonest parties lie in a square or rectangular section of the lattice; for example, where $D = \{ 8, 9, 10, 11, 12, 13, 14, 15, 16 \}$, or $D = \{ 5, 6, 7, 8, 9, 10, 11, 12, 13, 14, 15, 16 \}$.

We finish the proof by invoking Lemmas \ref{lem:fullstablem} and \ref{lem:finalbit}, taking $\mathsf{J}=n$.

\end{proof}

Note that there are many other sets of dishonest parties for the $\mathsf{2D}$ cluster state for which passing only the stabiliser generator measurements perfectly implies $\ket{\Psi} = U_D \ket{\mathscr{G}}$; as before, here we are only giving some examples of cases where such a simplification is allowed.
%

\newpage

\section{APPENDIX III: PROOF OF THEOREM \ref{th:graphdishonestverproof}}

In Protocol \ref{alg:dishonestver}, we outlined a symmetric protocol for verification of any graph state with a randomly chosen party playing the role of the Verifier. We will now prove its security in the following Theorem.

\begin{manualtheorem}{3}
If we set $N_{total} = (\lambda+1)\lambda \mathsf{J} N_{test}$ and $N_{test} = \lceil{ m \mathsf{J}^4 \ln \mathsf{J} }\rceil$, 
the probability that the fidelity of the averaged state of 
the target copy (over all possible choices of the tested copies and target copy)
in Protocol \ref{alg:dishonestver} satisfies
\begin{align}
F(\rho_H^{\ket{\mathscr{G}}}, \rho_H^{avg}) 
\geq 1 - \Big( \frac{1}{\lambda} - \frac{1}{\lambda^2} \Big) - \Big( 1 + \frac{1}{\lambda} \Big) \Big[ \frac{\sqrt{c}}{\mathsf{J}} + \lambda \mathsf{J} - \frac{N_{pass}}{N_{test}} \Big]
\end{align}
is at least $
\Big[ 1 - \sum_{x = 0}^{\lambda} \big( 1 - \frac{\abs{H}}{n} \big)^x \big( \frac{\abs{H}}{n} \mathsf{J}^{-\frac{2cm}{3}} \big)^{\lambda - x} \Big]^{\mathsf{J}},
$
where
$m, c$ are positive constants chosen such that the fidelity and probability expressions are greater than zero.
\end{manualtheorem}

\begin{proof}
Recall that by Serfling's bound, we have
\begin{align}
\text{Pr} \Big[ \sum_{t \in \overline{\mathsf{\Pi}} } Y_t \leq \frac{\mathcal{L}}{\mathcal{R}} \sum_{t \in \mathsf{\Pi}} Y_t + \mathcal{L}\nu \Big] 
\geq 1 - \exp \big[ - \frac{2 \nu^2 \mathcal{L} \mathcal{R}^2}{(\mathcal{L}+\mathcal{R})(\mathcal{R}+1)} \big].
\end{align}
In step 2 of Protocol \ref{alg:dishonestver}, each $\mathcal{S}_j$ is tested in $\lambda$ sets of $N_{test}$ copies, with a Verifier chosen at random for each set of $N_{test}$ copies.
Now consider the case where we measure the $l^{th}$ set of $N_{test}$ copies for the stabiliser $\mathcal{S}_j$.
This corresponds to  taking $\mathcal{L} = N_{total} - [(j-1) \lambda + l] N_{test}$ and $\mathcal{R} = N_{test}$.
Let $\mathsf{\Pi}^{(j,l)}$ be the $l^{th}$ set of copies on which each $S_j$ is measured, and $\overline{\mathsf{\Pi}}^{(j, l)}$
be the set of remaining copies after measuring. Applying Serfling's bound and assuming an honest Verifier, we get
\begin{align} \label{eq APP Serfling}
\text{Pr} \Big[ \sum_{t \in \overline{\mathsf{\Pi}}^{(j, l)} } Y_t \leq \frac{N_{total} - [(j-1)\lambda + l] N_{test}}{N_{test}} \sum_{t \in \mathsf{\Pi}^{(j, l)}} Y_t + \big(N_{total} - [(j-1)\lambda + l] N_{test} \big) \nu \Big]
\geq q_j^l.
\end{align}

The number of remaining copies after all tests is $\mathsf{N} = N_{total} - \lambda \mathsf{J} N_{test} = \lambda^2 \mathsf{J} N_{test}$. To bound $\mathsf{k}$, which is the number of `good' copies in the remaining, we must determine the number of `bad' copies for each stabiliser measurement $\mathcal{S}_j$.
Recall that for each $\lambda N_{test}$ copies that we test with stabiliser $\mathcal{S}_j$, the CRS randomly chooses a different Verifier to test each set of $N_{test}$ copies. 

Let us now assume that for each stabiliser test $\mathcal{S}_j$, at least one Verifier (out of $\lambda$) is honest, meaning that $\lambda - 1$ Verifiers are dishonest. For the $(\lambda - 1) N_{test}$ copies tested by a dishonest Verifier, we cannot conclude anything from our test, since the Verifier can let any state could pass the test. So, for all $\mathcal{S}_j$ copies, we have a total of $(\lambda -1) \mathsf{J} N_{test}$ copies that we will have to assume are bad. This means that we can lower bound $\mathsf{k}$ by 
\begin{align}
\mathsf{k} 
& \geq (N_{total} - \lambda \mathsf{J} N_{test}) - (\lambda -1) \mathsf{J} N_{test} - \sum_{j=1}^{\mathsf{J}} \Big[ \frac{N_{total} - [(j-1)\lambda] N_{test}}{N_{test}}  \sum_{t \in \mathsf{\Pi}^{(j, l)}} Y_t + \big( N_{total} - [(j-1)\lambda] N_{test} \big) \nu \Big] \nonumber \\
& = N_{total} - (2\lambda -1) \mathsf{J} N_{test} - \sum_{j=1}^{\mathsf{J}} \Big[  \frac{N_{total} - [(j-1)\lambda] N_{test}}{N_{test}}  \sum_{t \in \mathsf{\Pi}^{(j, l)}} Y_t + \big( N_{total} - [(j-1)\lambda] N_{test} \big) \nu \Big]  \nonumber \\
& = [(\lambda^2 + \lambda) - (2\lambda - 1)] \mathsf{J} N_{test} - \sum_{j=1}^{\mathsf{J}} \frac{N_{total}}{N_{test}} \sum_{t \in \mathsf{\Pi}^{(j, l)}} Y_t  
+ \sum_{j=1}^{\mathsf{J}} \frac{[(j-1)\lambda] N_{test}}{N_{test}} \sum_{t \in \mathsf{\Pi}^{(j, l)}} Y_t 
- \sum_{j=1}^{\mathsf{J}} N_{total} \nu + \sum_{j=1}^{\mathsf{J}} [(j-1)\lambda] N_{test} \nu  \nonumber \\
& \geq (\lambda^2 - \lambda + 1) \mathsf{J} N_{test} - \frac{N_{total}}{N_{test}} \sum_{j=1}^{\mathsf{J}} \sum_{t \in \mathsf{\Pi}^{(j, l)}} Y_t - \sum_{j=1}^{\mathsf{J}} N_{total} \nu  \nonumber \\
& = (\lambda^2 - \lambda +1) \mathsf{J} N_{test} - \frac{N_{total}}{N_{test}} \big( \lambda \mathsf{J} N_{test} - N_{pass} \big) - \mathsf{J} N_{total} \nu  \nonumber \\
& = (\lambda^2 - \lambda + 1) \mathsf{J} N_{test} - (\lambda^2 + \lambda) \mathsf{J} (\lambda \mathsf{J} N_{test} - N_{pass}) - (\lambda^2 + \lambda) \mathsf{J}^2 \nu N_{test}  \nonumber \\
& =  \Big[ (\lambda^2 - \lambda + 1) \mathsf{J} - (\lambda^2 + \lambda) \lambda \mathsf{J}^2 + (\lambda^2 + \lambda) \mathsf{J} \frac{N_{pass}}{N_{test}} - (\lambda^2 + \lambda) \mathsf{J}^2 \nu \Big] N_{test},
\end{align}
which gives the fraction of `good' remaining copies as 
\begin{align}
\frac{\mathsf{k}}{\mathsf{N}} 
& \geq \frac{ \Big[ (\lambda^2 - \lambda + 1) \mathsf{J} - (\lambda^2 + \lambda) \lambda \mathsf{J}^2 + (\lambda^2 + \lambda) \mathsf{J} \frac{N_{pass}}{N_{test}} - (\lambda^2 + \lambda) \mathsf{J}^2 \nu \Big] N_{test} }
{\lambda^2 \mathsf{J} N_{test}}  \nonumber \\
& = 1 - \Big( \frac{1}{\lambda} - \frac{1}{\lambda^2} \Big) - \Big( \frac{\lambda^3 + \lambda^2}{\lambda^2} \Big) \mathsf{J} + \Big( 1 + \frac{1}{\lambda} \Big) \frac{N_{pass}}{N_{test}} - \Big( 1 + \frac{1}{\lambda} \Big) \mathsf{J} \nu  \nonumber \\
& = 1 - \Big( \frac{1}{\lambda} - \frac{1}{\lambda^2} \Big) - \Big( 1 + \frac{1}{\lambda} \Big) \Big[ \frac{\sqrt{c}}{\mathsf{J}} + \lambda \mathsf{J} - \frac{N_{pass}}{N_{test}} \Big].
\end{align}
Recall that the fidelity $F(\rho_H^{\ket{\mathscr{G}}}, \rho_H^{avg}) \geq \frac{\mathsf{k}}{\mathsf{N}}$, and so we have
\begin{align} \label{eq APP Fidelity bound}
F(\rho_H^{\ket{\mathscr{G}}}, \rho_H^{avg}) \geq 1 - \Big( \frac{1}{\lambda} - \frac{1}{\lambda^2} \Big) - \Big( 1 + \frac{1}{\lambda} \Big) \Big[ \frac{\sqrt{c}}{\mathsf{J}} + \lambda \mathsf{J} - \frac{N_{pass}}{N_{test}} \Big].
\end{align}

Let us now calculate $q_j^l$. We get
\begin{align} \label{eq APP q_j^l}
q_j^l 
& = 1 - \exp \Big[ - \frac{2 \nu^2 \mathcal{L} \mathcal{R}^2}{(\mathcal{L}+\mathcal{R})(\mathcal{R}+1)} \Big]  \nonumber \\
& = 1 - \exp \Big[ - 2 \nu^2 \frac{\big( N_{total} - [(j-1)\lambda + l] N_{test} \big) N_{test}^2}{\big( N_{total} - [(j-1)\lambda + l] N_{test} + N_{test} \big) (N_{test} + 1)} \Big]  \nonumber \\
& = 1 - \exp \Big[ - 2 \nu^2 N_{test} \frac{1}{\frac{N_{total} - [(j-1)\lambda + l] N_{test} + N_{test} }{N_{total} - [(j-1)\lambda + l] N_{test} }} \frac{1}{\frac{N_{test} + 1}{N_{test}}} \Big]  \nonumber \\
& = 1 - \exp \Big[ - 2 \nu^2 N_{test} \frac{1}{1 + \frac{N_{test}}{N_{total} - [(j-1)\lambda + l] N_{test}}} \frac{1}{1 + \frac{1}{N_{test}}} \Big]  \nonumber \\
& = 1 - \exp \Big[ - 2 \nu^2 N_{test} \frac{1}{1 + \frac{1}{\lambda^2 \mathsf{J} + \lambda \mathsf{J} - \lambda j + \lambda - l}} \frac{1}{1 + \frac{1}{N_{test}}} \Big].
\end{align}
Setting $j = \mathsf{J}, l = \lambda$, we get
\begin{align} 
q_{\mathsf{J}} = 1 - \exp \Big[ - 2 \nu^2 N_{test} \frac{1}{1 + \frac{1}{\lambda^2 \mathsf{J}}} \frac{1}{1 + \frac{1}{N_{test}}} \Big].
\end{align}
Noting that $\lambda \geq 1, \mathsf{J} \geq 2, N_{test} \geq 1$, we get the probability that there are $\mathsf{k}$ `good' copies out of the remaining to be at least 
\begin{align}
q_{\mathsf{J}} \geq 1 - \exp \Big[ - \frac{2}{3} \nu^2 N_{test} \Big] = 1 - \mathsf{J}^{-\frac{2cm}{3}}.
\end{align}


We now calculate the probability that the above occurs, \emph{i.e.} that for each stabiliser test $\mathcal{S}_j$, one or more of the Verifiers are honest, and that when they are, the condition on the left-hand side of Equation (\ref{eq APP Serfling}) holds, leading to Equation (\ref{eq APP Fidelity bound}). 
For a given $\mathcal{S}_j$, this probability is given by
\begin{align}
1-    \sum_{x=0}^{\lambda} \left(1-\frac{|H|}{n}\right)^x \left(\frac{|H|}{n}\right)^{\lambda-x}\left(1-q_j^{l_{max}}\right)^{\lambda-x},
\end{align}
where $q_j^{l_{max}}$ is the maximum of $q_j^l$ over all $l$. From Equation 
(\ref{eq APP q_j^l}), we see that $q_j^l \geq q_\mathsf{J}$. Taking this for all $\mathcal{S}_j$, we get the probability that Equation (\ref{eq APP Fidelity bound}) holds to be 
\begin{align}
& \prod_{j=1}^{\mathsf{J}} \Big[ 1-    \sum_{x=0}^{\lambda} \left(1-\frac{|H|}{n}\right)^x \left(\frac{|H|}{n}\right)^{\lambda-x}\left(1-q_j^{l_{max}}\right)^{\lambda-x} \Big] \nonumber \\
& \geq \Big[ 1-    \sum_{x=0}^{\lambda} \left(1-\frac{|H|}{n}\right)^x \left(\frac{|H|}{n}\right)^{\lambda-x}\left(1-q_{\mathsf{J}} \right)^{\lambda-x} \Big]^{\mathsf{J}} \nonumber \\
& \geq \Big[ 1-    \sum_{x=0}^{\lambda} \left(1-\frac{|H|}{n}\right)^x \left(\frac{|H|}{n}\mathsf{J}^{-\frac{2cm}{3}}\right)^{\lambda-x}\Big]^\mathsf{J}.
\end{align}


\end{proof}

\section{APPENDIX IV: ROBUSTNESS}

When considering the robustness of our protocols, one is interested in how the security and probability of accept depend on the noise that is introduced (eg. as in  \cite{Takeuchi2019,unnikrishnan2019authenticated}).
For example, if we satisfy the condition in step 4 of Protocols \ref{alg:takeuchi} and \ref{alg:dishonestver}, \emph{i.e.} we have \begin{align}
    N_{pass} \equiv \sum_{j = 1}^{\mathsf{J}} N_{pass, j} \geq \lambda \mathsf{J} N_{test} - \frac{N_{test}}{2  \mathsf{J}},
    \end{align}
then Theorems \ref{th:graphbigproof}, \ref{th:allexamples} and \ref{th:graphdishonestverproof} give fidelity
\begin{align}
F(\rho_H^{\ket{\mathscr{G}}}, \rho_H^{avg}) 
 \geq  1 - \Big( \frac{1}{\lambda} - \frac{1}{\lambda^2} \Big)  - \Big( 1 + \frac{1}{\lambda} \Big) \Big[ \frac{2\sqrt{c}+1}{\mathsf{2J}}  \Big],
\end{align}
where we take $\lambda = 1$ for the cases in Theorems \ref{th:graphbigproof} and \ref{th:allexamples}, $\mathsf{J} = 2^n$ for the general case (Theorems \ref{th:graphbigproof} and \ref{th:graphdishonestverproof}) and $\mathsf{J} = n$ when generator tests are enough (Theorems \ref{th:allexamples} and \ref{th:graphdishonestverproof}).

Following \cite{Takeuchi2019}, let us take the example of the following noisy state being provided by the source
\begin{align}
 \rho = \sigma^{\otimes N_{total}},   
\end{align}
where $\sigma=(1-\epsilon) |G\rangle\langle G| + \epsilon \eta$, for some noise state $\eta$ and noise parameter $\epsilon$. Then, the probability of the state being used for an application (step 4) is bounded by
\begin{align}
    p_{app} \geq \sum_{k=1}^{N_{test}/2\mathsf{J}} \binom{\lambda \mathsf{J} N_{test}}{k}\left(1-\epsilon\right)^{\lambda \mathsf{J}  N_{test}-k}\epsilon^k,
\end{align}
which goes to one with $n$ when $\epsilon < O\left(\frac{1}{\lambda \mathsf{J} N_{test}}\right)$.
This has the same dependency on the number of tests as \cite{Takeuchi2019} when $\lambda=1$. 
The cost of the symmetric version scales only linearly with $\lambda$; however, the cost with the number of parties for the general graph state is exponentially higher once more. Thus, for the general case the scheme loses out on robustness, but this is recovered for cases such as those in Theorem \ref{th:allexamples} where the generator tests suffice.

\end{document}